\documentclass[longauth]{aa} % for the long lists of affiliations
\usepackage{graphicx}
\usepackage[varg]{txfonts}
\usepackage{color}
\usepackage{natbib}
\bibpunct{(}{)}{;}{a}{}{,} % to follow the A&A style
\def\mpcoh{\,h^{-1}{\rm Mpc}}
\def\kms{\,{\rm km\, s}^{-1}}

\def\rmops{R\textsc{mocks }}
\def\vmops{V\textsc{mocks }}
\def\rmos{R\textsc{mock }}
\def\vmos{V\textsc{mock }}
\def\rmop{R\textsc{mocks}}
\def\vmop{V\textsc{mocks}}

\def\vmo{V\textsc{mock}}

\usepackage{amstext}

\begin{document}

  \title{The VIMOS Public Extragalactic Redshift Survey
    (VIPERS). \thanks{based on observations collected at the European
      Southern Observatory, Cerro Paranal, Chile, using the Very Large
      Telescope under programs 182.A-0886 and partly 070.A-9007.  Also
      based on observations obtained with MegaPrime/MegaCam, a joint
      project of CFHT and CEA/DAPNIA, at the Canada-France-Hawaii
      Telescope (CFHT), which is operated by the National Research
      Council (NRC) of Canada, the Institut National des Sciences de
      l’Univers of the Centre National de la Recherche Scientifique
      (CNRS) of France, and the University of Hawaii. This work is
      based in part on data products produced at TERAPIX and the
      Canadian Astronomy Data Centre as part of the
      Canada-France-Hawaii Telescope Legacy Survey, a collaborative
      project of NRC and CNRS. The VIPERS web site is
      http://www.vipers.inaf.it/. }} \subtitle{The decline of cosmic star formation: quenching,  mass, and environment connections}

\titlerunning{Quenching,  mass, and environment in VIPERS}

   \author{O.~Cucciati\inst{\ref{oabo},\ref{unibo}}           
\and I.~Davidzon\inst{\ref{lam},\ref{oabo}}   
\and M.~Bolzonella\inst{\ref{oabo}}      
\and B.~R.~Granett\inst{\ref{brera},\ref{unimi}}                                
\and G.~De Lucia\inst{\ref{oats}}
\and E.~Branchini\inst{\ref{roma3},\ref{infn-roma3},\ref{oa-roma}}
\and G.~Zamorani\inst{\ref{oabo}}
\and A.~Iovino\inst{\ref{brera}}
\and B.~Garilli\inst{\ref{iasf-mi}}
\and L.~Guzzo\inst{\ref{brera},\ref{unimi}}
\and M.~Scodeggio\inst{\ref{iasf-mi}}
\and S.~de la Torre\inst{\ref{lam}}
\and U.~Abbas\inst{\ref{oa-to}}
\and C.~Adami\inst{\ref{lam}}
\and S.~Arnouts\inst{\ref{lam}}
\and D.~Bottini\inst{\ref{iasf-mi}}
\and A.~Cappi\inst{\ref{oabo},\ref{nice}}
\and P.~Franzetti\inst{\ref{iasf-mi}}   
\and A.~Fritz\inst{\ref{iasf-mi}}       
\and J.~Krywult\inst{\ref{kielce}}
\and V.~Le Brun\inst{\ref{lam}}
\and O.~Le F\`evre\inst{\ref{lam}}
\and D.~Maccagni\inst{\ref{iasf-mi}}
\and K.~Ma{\l}ek\inst{\ref{warsaw-nucl}}
\and F.~Marulli\inst{\ref{unibo},\ref{infn-bo},\ref{oabo}} 
\and T.~Moutard\inst{\ref{halifax},\ref{lam}}  
\and M.~Polletta\inst{\ref{iasf-mi},\ref{marseille-uni},\ref{toulouse}}
\and A.~Pollo\inst{\ref{warsaw-nucl},\ref{krakow}}
\and L.A.M.~Tasca\inst{\ref{lam}}
\and R.~Tojeiro\inst{\ref{st-andrews}}  
\and D.~Vergani\inst{\ref{iasf-bo}}
\and A.~Zanichelli\inst{\ref{ira-bo}}
\and J.~Bel\inst{\ref{cpt}}
\and J.~Blaizot\inst{\ref{lyon}}
\and J.~Coupon\inst{\ref{geneva}}
\and A.~Hawken\inst{\ref{brera},\ref{unimi}}
\and O.~Ilbert\inst{\ref{lam}}
\and L.~Moscardini\inst{\ref{unibo},\ref{infn-bo},\ref{oabo}}
\and J.~A.~Peacock\inst{\ref{roe}}
\and A.~Gargiulo\inst{\ref{iasf-mi}}
}

%   \offprints{Olga Cucciati \\ \email{olga.cucciati@oabo.inaf.it}}

\institute{
INAF - Osservatorio Astronomico di Bologna, via Ranzani 1, I-40127, Bologna, Italy \\ \email{olga.cucciati@oabo.inaf.it} \label{oabo} %4
\and Dipartimento di Fisica e Astronomia - Alma Mater Studiorum Universit\`{a} di Bologna, viale Berti Pichat 6/2, I-40127 Bologna, Italy \label{unibo}%17
\and Aix Marseille Univ, CNRS, LAM, Laboratoire d'Astrophysique de
Marseille, Marseille, France  \label{lam}%5
\and INAF - Osservatorio Astronomico di Brera, Via Brera 28, 20122 Milano
--  via E. Bianchi 46, 23807 Merate, Italy \label{brera}%1
\and  Universit\`{a} degli Studi di Milano, via G. Celoria 16, 20133 Milano, Italy \label{unimi}%2
\and INAF - Osservatorio Astronomico di Trieste, via G. B. Tiepolo 11, 34143 Trieste, Italy \label{oats}%13
\and Dipartimento di Matematica e Fisica, Universit\`{a} degli Studi Roma Tre, via della Vasca Navale 84, 00146 Roma, Italy\label{roma3} %10
\and INFN, Sezione di Roma Tre, via della Vasca Navale 84, I-00146 Roma, Italy \label{infn-roma3}%28
\and INAF - Osservatorio Astronomico di Roma, via Frascati 33, I-00040 Monte Porzio Catone (RM), Italy \label{oa-roma}%29
\and INAF - Istituto di Astrofisica Spaziale e Fisica Cosmica Milano, via Bassini 15, 20133 Milano, Italy \label{iasf-mi}%3
\and INAF - Osservatorio Astrofisico di Torino, 10025 Pino Torinese, Italy \label{oa-to}%5
\and Laboratoire Lagrange, UMR7293, Universit\'e de Nice Sophia Antipolis, CNRS, Observatoire de la C\^ote d'Azur, 06300 Nice, France \label{nice}%
\and Institute of Physics, Jan Kochanowski University, ul. Swietokrzyska 15, 25-406 Kielce, Poland \label{kielce}%15
\and National Centre for Nuclear Research, ul. Hoza 69, 00-681 Warszawa, Poland \label{warsaw-nucl}%23
\and INFN, Sezione di Bologna, viale Berti Pichat 6/2, I-40127 Bologna, Italy \label{infn-bo}%18
\and Department of Astronomy \& Physics, Saint Mary's University, 923 Robie Street, Halifax, Nova Scotia, B3H 3C3, Canada \label{halifax}%13
\and Aix-Marseille Universite, Jardin du Pharo, 58 bd Charles Livon, F-13284 Marseille cedex 7, France \label{marseille-uni}
\and IRAP,  9 av. du colonel Roche, BP 44346, F-31028 Toulouse cedex 4, France \label{toulouse} 
\and Astronomical Observatory of the Jagiellonian University, Orla 171, 30-001 Cracow, Poland \label{krakow} %22
\and School of Physics and Astronomy, University of St Andrews, St Andrews KY16 9SS, UK \label{st-andrews}%11
\and INAF - Istituto di Astrofisica Spaziale e Fisica Cosmica Bologna, via Gobetti 101, I-40129 Bologna, Italy \label{iasf-bo}%25
\and INAF - Istituto di Radioastronomia, via Gobetti 101, I-40129,
Bologna, Italy \label{ira-bo}%26
\and Aix Marseille Univ, Univ Toulon, CNRS, CPT, Marseille, France \label{cpt}%7
\and Univ. Lyon, Univ. Lyon1, ENS de Lyon, CNRS, Centre de Recherche Astrophysique de Lyon, UMR5574, F-69230, Saint-Genis-Laval, France \label{lyon}%8
\and Department of Astronomy, University of Geneva, ch. d'Ecogia 16, 1290 Versoix, Switzerland \label{geneva}%12
\and Institute for Astronomy, University of Edinburgh, Royal
Observatory, Blackford Hill, Edinburgh EH9 3HJ, UK \label{roe}%14
}

% \abstract{}{}{}{}{} 
% 5 {} token are mandatory

\abstract{We use the final data of the VIMOS Public Extragalactic
  Redshift Survey (VIPERS) to investigate the effect of the
  environment on the evolution of galaxies between $z=0.5$ and
  $z=0.9$.  We characterise local environment in terms of the density
  contrast smoothed over a cylindrical kernel, the scale of which is
  defined by the distance to the fifth nearest neighbour. This is
  performed by using a volume-limited sub-sample of galaxies complete
  up to $z=0.9$, but allows us to attach a value of local density to
  all galaxies in the full VIPERS magnitude-limited sample to
  $i<22.5$. We use this information to estimate how the distribution
  of galaxy stellar masses depends on environment. More massive
  galaxies tend to reside in higher-density environments over the full
  redshift range explored. Defining star-forming and passive galaxies
  through their (NUV$-r$) vs ($r-K$) colours, we then quantify the
  fraction of star-forming over passive galaxies, $f_{\rm ap}$, as a
  function of environment at fixed stellar mass. $f_{\rm ap}$ is
  higher in low-density regions for galaxies with masses ranging from
  $\log(\mathcal{M}/\mathcal{M}_\odot)=10.38$ (the lowest value
  explored) to at least $\log(\mathcal{M}/\mathcal{M}_\odot)\sim11.3$,
  although with decreasing significance going from lower to higher
  masses. This is the first time that environmental effects on
  high-mass galaxies are clearly detected at redshifts as high as
  $z\sim0.9$.  We compared these results to VIPERS-like galaxy mock
  catalogues based on a widely used galaxy formation model.  The model
  correctly reproduces $f_{\rm ap}$ in low-density environments, but
  underpredicts it at high densities.  The discrepancy is particularly
  strong for the lowest-mass bins. We find that this discrepancy is
  driven by an excess of low-mass passive satellite galaxies in the
  model. In high-density regions, we obtain a better (although not
  perfect) agreement of the model $f_{\rm ap}$ with observations by
  studying the accretion history of these model galaxies (that is, the
  times when they become satellites),  by
  assuming either that a non-negligible fraction of satellites is
  destroyed, or that their quenching timescale is longer than $\sim 2$
  Gyr.  }

%{}{}{}{}

% \abstract{}{}{}{}{} 
% 5 {} token are mandatory
 %\abstract
%   {Context}
%   {Aims}
%   {Methods}
%   {Results}
%   {Conclusions}


   \maketitle

%****************************************************************************
\section{Introduction}\label{intro}

Since pioneering work about four decades ago
\citep[e.g.][]{oemler74,davis_geller76,dressler80}, environmental
studies have increased in importance in the context of galaxy
evolution.  The first observations found two distinct galaxy
populations (red and elliptical  vs blue and spiral) residing in different
environments in the local Universe. More recent surveys extended this
fundamental result to higher redshifts
\citep[e.g.][]{cucciati2006,cooper07}, and/or replaced the
visual or colour classification with estimates of the star formation rate
(SFR) or other indicators of the dominant stellar population such as
the measurement of the D4000\AA\ break \citep[see
e.g.][]{balogh98,hashimoto98,gomez03,kauffmann04,grutzbauch11b}.

The environment reconstruction has also been improved over the
years. The systematic identification of galaxy clusters and groups
allowed the community to perform more detailed analysis of galaxy
populations in different environments \citep[see e.g.][and references
therein]{cucciati10,iovino10,
  gerke12,knobel13,kovac14,annunziatella14,haines15}. Furthermore, the
development of new methods to compute the local density around
galaxies (such as Voronoi tessellation) have enabled both the
identification of galaxy clusters and the parameterisation of the
density field as a whole to become more reliable 
\citep{marinoni02,cooper05,kovac2010_density,lemaux16,fossati16}. Moreover, the
complex topology of the large-scale structure (LSS) can now be
dissected, spanning from the large-scale filamentary cosmic web
\citep[e.g.][]{tempel13,einasto14,alpaslan14,malavasi17} to detailed
analysis focused on smaller regions \citep[e.g. single clusters or
walls, as in][]{gavazzi10,boselli14,iovino16}.

In this context, spectroscopic galaxy surveys play a pivotal role in
identifying LSS both on small and large scales. Several analyses have
used some of the most precise photometric redshifts available to date
\citep{scoville13,darvish15,malavasi16}. Despite this, spectroscopic
measurements of galaxy redshifts ($z_\mathrm{spec}$) are generally
required in order to minimise the uncertainties in the radial position
of galaxies. In some cases, the $z_\mathrm{spec}$ measurement error is
so small that the strongest limitation is due to peculiar velocities
\citep{kaiser87}.  Large and deep spectroscopic surveys comprise the
best data-sets to study how environment affects galaxy evolution,
thanks to their large volume and the long time-span covered. They are,
however, very time consuming to assemble.

At present, only the VIMOS Public Extragalactic Redshift Survey
\citep[VIPERS,][]{guzzo14} offers the desired combination of large
volume ($5 \times 10^7\, h^{-3}$~Mpc$^3$) and precise galaxy redshifts
at $z>0.5$.  VIPERS was conceived as a high-redshift ($0.5<z<1.2$)
analogue of large local surveys like 2dFGRS \citep[][]{colless01}.
With respect to other surveys at intermediate redshifts --
for example,~zCOSMOS \citep{lilly2009_zCOSMOS}, which has the same depth as
VIPERS -- the larger volume covered by VIPERS significantly reduces
the effect of cosmic variance \citep[which has important effects in
zCOSMOS: see e.g.][]{delatorre10}. This allows us to study rare galaxy
populations, such as the most massive galaxies, with more solidity \citep[see][]{davidzon13}.

Of course we also need an interpretative architecture in which to
frame our observations. Fortunately, today we have
sophisticated simulations and theoretical models of galaxy formation
and evolution at our disposal that can help us in this task, together with
simulations of dark matter (DM) halo merger trees. With respect to the
observations, these theoretical tools offer us the advantage to study
the relationship between baryonic and dark matter, to link galaxy
populations at different redshifts (e.g. the problem of finding the
progenitors of a given galaxy population), and study
the environmental history of galaxies in detail
\citep[e.g.][]{gabor10,delucia12,hirschmann14}.

Much progress has been made with simulations in recent years, with
larger simulated boxes (see e.g. the Bolshoi simulation,
\citealp{klypin11}, and the MultiDark run,
\citealp{prada2012_multidark}), better spatial resolution
\citep[e.g. the Millennium II simulation, ][]{boylan_kolchin09}, and
the implementation of hydrodynamical codes on cosmological volumes
(e.g. the EAGLE simulation, \citealp{schaye15}, and the ILLUSTRIS
simulation, \citealp{vogelsberger14}).  Much effort has also
been made to improve semi-analytical models of galaxy formation and
evolution \citep[see
e.g.][]{guo11,delucia14,henriques15,hirschmann16_GAEA}. Although
  several works have studied the role of the environment in models of
  galaxy evolution \citep[see
  e.g.][]{cen11,delucia12,hirschmann14,henriques16},  some
limitations still remain, such as the environment definition, which
has to be linked to observational quantities in order to make a
meaningful comparison between the models and real data
\citep[see][]{muldrew12,haas12,hirschmann14,fossati15}.

As a final note, we remark that the way in which we ask ourselves
the questions to be answered has also evolved in recent years.  As an
example, the wide-spread scenario of `nature vs nurture' in galaxy
evolution has been questioned, and it might well be an ill-posed
problem.  In fact, even if we possessed an ideal set of simulations
and observations, it would be misleading to analyse them by
contrasting environmental effects with the evolution driven by
intrinsic galaxy properties (such as the stellar or halo mass). These
two aspects are physically connected, and it is impossible to fully
separate them \citep[see the discussion in][]{delucia12}.

With this picture in mind, we aim at using VIPERS to shed new light on
galaxy evolution and environment.  In another paper of this series
\citep{malavasi17} we show a reconstruction of the cosmic web, while
in this paper we present the density field of the final VIPERS sample.
Our goal is to study how environment affects the evolution of the
galaxy specific star formation rate (sSFR) and compare it with
simulations to obtain new insights into the mechanisms that halt star
formation (i.e., `quenching').  The paper is organised as follows.  In
Sect.~\ref{data_mocks} we briefly describe the VIPERS sample and the
mock galaxy catalogues we use in our analysis. In
Sect.~\ref{density_field} we present the VIPERS density field, and we
show how environment affects galaxy stellar mass and sSFR in
Sect.~\ref{SM_sSFR_env}. In Sect.~\ref{simul} we compare our results
to a similar analysis performed in the mock galaxy catalogues. We
discuss our findings in Sect.~\ref{discussion} and summarise our work
in Sect.~\ref{summary}. In the Appendices we give additional details
on the reliability of the density field reconstruction, and we show how
the final VIPERS density field compares to that reconstructed from
VIPERS first data release.

Except where explicitly stated, we assume a flat $\Lambda$CDM
cosmology throughout the paper with $\Omega_m=0.30$,
$\Omega_{\Lambda}=0.70$, $H_0=70\kms {\rm Mpc}^{-1}$ and
$h=H_0/100$. Magnitudes are expressed in the AB system
\citep{oke74,fukugita96}.

%****************************************************************************

\section{Data and mock samples}\label{data_mocks}

\subsection{Data}\label{data}

VIPERS\footnote{http://vipers.inaf.it} \citep{guzzo14,scodeggio16} has
measured redshifts for $\sim 10^5$ galaxies at redshift $0.5 < z
\lesssim 1.2$.  The project had two broad scientific goals: i) to
reliably measure galaxy clustering and the growth of structure through
redshift-space distortions, ii) to study galaxy properties at an epoch
when the Universe was about half its current age, over a volume
comparable to that of large existing local ($z\sim0.1$) surveys, like
2dFGRS and SDSS.

The VIPERS global footprint covers a total of 23.5 deg$^2$, split over
the W1 and W4 fields of the Canada-France-Hawaii Telescope Legacy
Survey (CFHTLS) Wide. Targets were selected to $i_{\rm AB}<22.5$ from
the fifth data release (T0005, \citealp{CFHTLS}).  A colour
pre-selection in $(r-i)$ vs $(u-g)$ was also applied to remove
galaxies at $z<0.5$. Together with an optimised slit configuration
\citep{scodeggio2009_VIMOS}, this allowed us to obtain a target
sampling rate of $\sim 47\%$ over the redshift range of interest,
about doubling what we would have achieved by selecting a purely
magnitude-limited sample to the same surface density.

The VIPERS spectroscopic observations were carried out using the
VIsible Multi-Object Spectrograph (VIMOS,
\citealp{lefevre2002,lefevre2003}), using the low-resolution Red grism
($R\simeq220$ over the wavelength range 5500-9500$\AA$).  The number
of slits in each VIMOS pointing was maximised using the SSPOC
algorithm \citep{bottini2005}. The typical radial velocity error on
the spectroscopic redshift ($z_s$) measurement of a galaxy is
$\sigma_z=0.00054(1+z)$ (see \citealp{scodeggio16} for more
details). A discussion of the survey data reduction and database
system is presented in \citet{garilli2012}.

The data used here correspond to the publicly released PDR-2 catalogue
\citep{scodeggio16}, with the exception of a small sub-set of
redshifts (340 galaxies missing in the range $0.6 < z< 1.1$), for
which the redshift and quality flags were revised closer to the
release date. Concerning the analysis presented here, this has no
effect. We retain only galaxies with reliable redshift measurements,
defined as having quality flag equal to 2, 3, 4, and 9. The quality
flag is assigned to each targeted object during the process of
validating redshift measurements, according to a scheme that
has been adopted by previous VIMOS surveys (VVDS, \citealp{lefevre2005}, and
zCOSMOS, \citealp{lilly2009_zCOSMOS}).  The average confidence level
of single redshift measurements for the sample of reliable redshifts
is estimated to be 96.1\% \citep{scodeggio16}. In our case, this
selection produces a sample of 74835 galaxies.

We computed the survey selection function and assigned a set of three
weights to each galaxy with a reliable redshift: the colour sampling
rate (CSR), the target sampling rate (TSR), and the spectroscopic
success rate (SSR). The CSR takes into account the modification
of the redshift
distribution, $n(z)$, of a purely flux limited catalogue ($i_{\rm
  AB}<22.5$) by the colour pre-selection applied to remove
galaxies at $z<0.5$ from the sample. As a consequence, the CSR depends
on redshift, and so it smoothly varies from 0 to 1 from
$z\sim0.4$ to $z\sim0.6$, and it remains equal to 1 for $z\geq 0.6$.  The
TSR is the fraction of galaxies in the parent photometric catalogue
($i_{\rm AB}<22.5$ and colour cut) that have a slit placed over
them. Finally, the SSR is the fraction of targeted galaxies for which
a reliable redshift has been measured. Considering the TSR and SSR
together, VIPERS has an average effective sampling rate of $\sim40$\%.
In our computation, the TSR depends on the local projected density
around each target, while the SSR depends on $i-$band magnitude,
redshift, rest-frame colour, $B$-band luminosity, and the quality of
the VIMOS quadrants.

Unless otherwise specified, we use the W1 and W4 samples together as
the `VIPERS sample' throughout this paper. We refer to the sample of
galaxies with a reliable spectroscopic redshift as defined above as
`spectroscopic galaxies', and to all the other galaxies with only a
photometric redshift and with $i<22.5$ as `photometric galaxies'. We
refer to the entire flux limited catalogue limited at $i_{\rm
  AB}<22.5$, before the colour pre-selection, as the `parent
photometric catalogue'.

\subsection{Photometric redshifts, luminosities, and stellar
  masses}\label{SED}

As part of the VIPERS Multi-Lambda Survey
(VIPERS-MLS\footnote{http://cesam.lam.fr/vipers-mls/}, see
\citealp{moutard16a} for further details), photometry from the final
CFHTLS\footnote{http://www.cfht.hawaii.edu/Science/CFHTLS/} release
(T0007\footnote{http://terapix.iap.fr/cplt/T0007/doc/T0007-doc.html})
in the $ugriz$ filters was optimised to provide both accurate colours
and reliable pseudo-total magnitudes. From this photometry,
photometric redshifts ($z_p$) were computed for all galaxies in the
VIPERS photometric catalogue.  Far-UV (FUV) and near-UV (NUV) from GALEX
\citep{martin05_galex}, $ZYJHK$ filters from VISTA \citep{emerson04},
and $K_s$ from WIRCam \citep{puget04_WIRCam} were also used, when
available. $ZYJHK$ observations are part of VIDEO
\citep{jarvis13_VIDEO}.  Down to $i_{\rm AB}<22.5$, the photometric
redshift error is $\sigma_{zp} = 0.035(1+z)$, with a $<2\%$ of
outliers rate (see Fig.~12 in \citealp{moutard16a}).

Absolute magnitudes, stellar masses, and SFR were obtained through a
spectral energy distribution (SED) fitting technique, using the code
{\it Le
  Phare\footnote{http://www.cfht.hawaii.edu/$\sim$arnouts/LEPHARE/lephare.html}}
as in \citet[][M16b from now on]{moutard16b}.  The SED fitting
used all the photometric bands described above.

We used the stellar population synthesis models of \citet{BC03}, with
two metallicities ($Z=0.008$ and $Z=0.02$) and exponentially declining
star formation histories, defined by SFR $\propto e^{-t/\tau}$, with
SFR being the instantaneous star formation rate and nine different values
for $\tau$, ranging between 0.1 Gyr and 30 Gyr as in \citet{ilbert13}.
We adopted three extinction laws (\citealp{prevot84},
\citealp{calzetti2000}, and an intermediate-extinction curve as in
\citealp{arnouts13}). We imposed a maximum dust reddening of $E(B-V)$
$\leq$ 0.5 for all galaxies and a low extinction for low-SFR galaxies
($E(B-V)\leq$ 0.15 if age/$\tau$ > 4).  We took into account the
emission-line contribution as described in \cite{ilbert09}.  To
compute the absolute magnitudes, we minimised their dependency on the
template library by using the observed magnitude in the band closest
to the redshifted absolute magnitude filter, unless the closest
apparent magnitude had an error $> 0.3$ mag. We refer to
Appendix A.1 of \citet{ilbert05} for more details. The SFR assigned to
each galaxy is the instantaneous SFR (see above) of the best-fit
template at the redshift of the galaxy, and it is not constrained by
any prior. From the stellar mass $\mathcal{M}$ and the SFR we also
derived the specific SFR (sSFR), defined as
sSFR$=$SFR$/\mathcal{M}$. We computed absolute magnitudes and stellar
masses with the same method for both the spectroscopic and photometric
galaxies, using their $z_s$ and $z_p$, respectively.

\subsection{Mock samples}\label{mocks}

We make use of mock galaxy catalogues to test the reliability of the
density field reconstruction, and to investigate the physical
processes taking place in different environments by comparing how
environment affects galaxy evolution in the model and in the data.

Our mock galaxy catalogues were obtained by embedding the
semi-analytical model (SAM) of galaxy evolution described in
\cite{delucia_blaizot2007} within DM halo merging trees extracted from
the Millennium Simulation \citep{springel2005_MILL}. The mass of the
DM particles is $8.6 \times 10^8 h^{-1} M_ {\odot}$. The DM run
adopted a $\Lambda$CDM cosmology with $\Omega_m = 0.25$, $\Omega_b =
0.045$, $h = 0.73$, $\Omega_{\Lambda}= 0.75$, $ n = 1$, and $\sigma_8
= 0.9$. These SAM mock catalogues contain, among other galaxy
properties, the right ascension, declination, redshift (including
peculiar velocity), $i$-band observed magnitude, $B$-band absolute
magnitude, galaxy stellar mass, and SFR.  We remark that this cosmology
is outdated, with for instance  $\sigma_8$ based on the first-year results of
the {\it Wilkinson} Microwave Anisotropy Probe (WMAP1,
\citealp{spergel03_WMAP1}) being larger than more recent measurements
such as WMAP7 \citep{komatsu11_WMAP7} and WMAP9
\citep{hinshaw13_WMAP9}, where they find $\sigma_8$ to be of the order of
0.8. \cite{davidzon16} showed that the density distribution
is slightly different in two simulations based on the cosmologies from
WMAP1 and WMAP3, but WMAP3 had a very low $\sigma_8$
($\sigma_8$=0.7). Given that $\sigma_8$ (and also other cosmological
parameters) from WMAP7 and WMAP9 is closer to that of WMAP1, we do not
expect a large difference between simulations based on WMAP1 or WMAP7
and WMAP9, as also shown in \cite{guo2013_millWMAP7}.

We used 50 pseudo-independent light cones, each covering an area
corresponding to the VIPERS W4 field, from which we built mock galaxy
catalogues, as follows.

\begin{itemize}
\item[-] First, from each light cone we extracted a purely flux-limited
  catalogue with the same magnitude cut as VIPERS ($i\leq22.5$). We
  refer to these catalogues as `reference mock catalogues'
  (\rmops from now on). In the \rmops we retained the apparent
  redshift (cosmological redshift plus peculiar velocity) without
  adding any redshift measurement error. The density contrast computed
  with these catalogues ($\delta^R$) is the standard on which we
  assess how well we can measure $\delta$ in a VIPERS-like survey.
\item[-] Second, from each \rmos we built two catalogues: a
  VIPERS-like photometric catalogue, and a VIPERS-like spectroscopic
  catalogue. The photometric catalogue was obtained by mimicking the
  VIPERS photometric redshift measurement error by adding a random
  value extracted from a Gaussian distribution with $\sigma_{zp} =
  0.035(1+z)$ to the apparent redshift of the \rmos. The spectroscopic
  catalogue was obtained from the \rmos first by modelling the $n(z)$
  at $z<0.6$ to mimic the VIPERS CSR, then by applying the same slit-positioning software (SSPOC, see \citealp{bottini2005}) as was used to
  select VIPERS targets. In this way, we were able to obtain the same
  VIMOS footprint as in VIPERS (see Fig.~\ref{fields_fig}) and a TSR
  that varied between quadrants as in VIPERS. We did not model the SSR in
  the mock catalogue because it depends on a large variety of
  factors, such as redshift and magnitude. Nevertheless, to account
  for its net effect of reducing the final number of measured
  spectroscopic redshifts, we randomly removed some of the galaxies
  left after applying SSPOC, in order to reach the same average SSR as
  the VIPERS data. Finally, we mimicked the VIPERS spectroscopic
  redshift error by adding a random value extracted from a Gaussian
  distribution with $\sigma_{z} = 0.00054(1+z)$ to the apparent
  redshift. We refer to these photometric and spectroscopic mock
  catalogues as `VIPERS-like mocks' (\vmops from now on).
\end{itemize}

%****************************************************************************

\begin{figure*} \centering
\includegraphics[width=17.3cm]{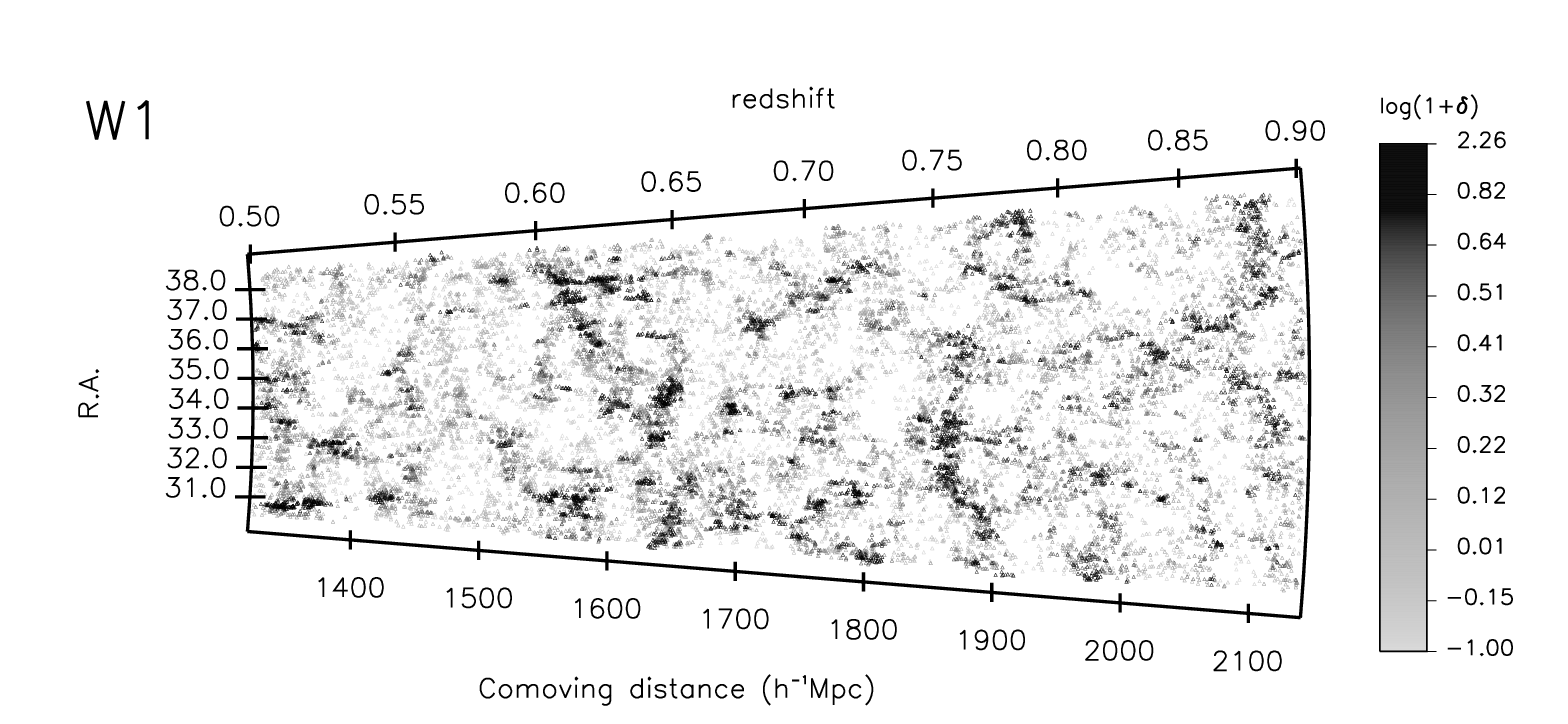}
\includegraphics[width=17.3cm]{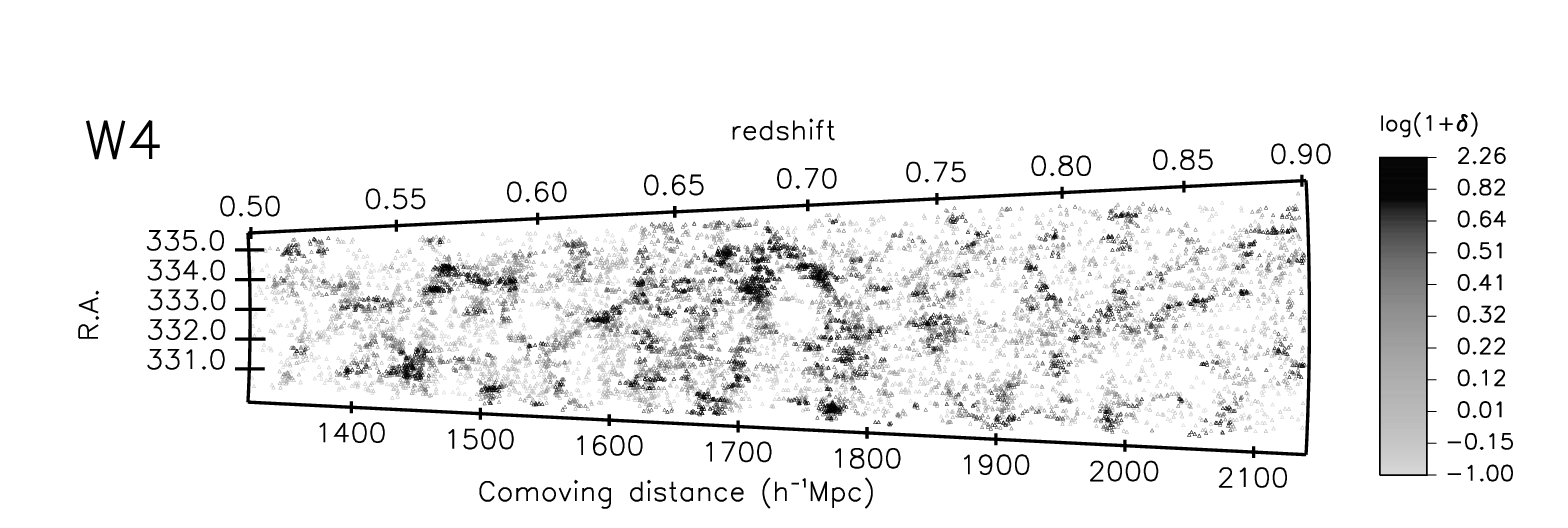}
\caption{${\rm R.A.}- z$ distribution of secure-redshift galaxies in W1
  (top) and W4 (below), in comoving coordinates. For the sake of
  clarity, only the central degree in Dec is plotted. The colour
  used for each galaxy refers to the value of the local density
  computed around the galaxy (from light grey for the lowest
density to
  black for the highest density, as in the colour bar). The density is
  computed in cylindrical filters with radius corresponding to the fifth n.n., using the volume-limited sample of
  tracers that is complete up to $z=0.9$.}
\label{density_fig} 
\end{figure*}

\section{VIPERS density field}\label{density_field}

We parameterised the local environment around each galaxy using the density
contrast $\delta,$ which is defined as 
\begin{equation} \displaystyle 
 \delta({\bf r}) \equiv \frac{\rho({\bf r}) - \langle \rho({\bf r}(z)) \rangle}{\langle \rho({\bf r}(z)) \rangle}, 
\label{delta_eq} 
\end{equation}
where $\rho({\bf r})$ is the local density at the comoving position
{\bf r} of each galaxy and $\langle \rho({\bf r}(z)) \rangle$ is the mean density at
that redshift.  We estimate $\rho({\bf r})$ using
counts-in-cells, as follows:
\begin{equation} \displaystyle 
 \rho({\bf r}) = \sum_{i}\frac{F({\bf r},R)}{\phi_i\,(m,z,{\rm R.A.},{\rm Dec}...)}. 
\label{rho_eq} 
\end{equation}

In Eq.~\ref{rho_eq} the sum runs over all the galaxies of the sample
used to trace the density field. We call these galaxies
`tracers'. $F({\bf r},R)$ is the smoothing filter (with scale $R$)
over which the density is computed, and $\phi$ is the selection
function of the sample. We always work in redshift space. In this work
we use the fractional density perturbation $\delta$, but we often
refer to it simply as `density' for the sake of simplicity.

The computation of $\rho$ depends on a variety of options regarding
the filter shape, the sample of galaxies to be used as tracers, how to
take into account the spectroscopic sampling rate, etc. These choices
are normally the result of a compromise between the characteristics of
the survey and the scientific goal. We refer to \cite{kovac2010_density},
for example, for an extensive discussion of these
alternatives. See Appendix \ref{app_density} for a detailed discussion
of our specific calculation for VIPERS and of the tests we have made
to quantify the reliability of our calculation.

We use the density field computed with cylindrical top-hat filters with a half-length of $1000\kms$ and radius corresponding to
the distance to the fifth nearest neighbour (`n.n.' from now on),
using a volume-limited sample of tracers with a luminosity cut given
by $M_B \leq -20.4-z$. This luminosity limit makes the tracer sample
complete up to $z=0.9,$ and (more importantly) this selection
empirically yields a comoving number density that does not evolve, so
that the meaning of our densities is not affected by discreteness
effects that change with redshift.  Cylinders are centred around all
of the galaxies in our sample. With this cylindrical filter, both
$\rho({\bf r})$ and $\langle \rho({\bf r}(z)) \rangle$ have the
dimensions of surface densities in redshift slices of $\pm 1000\kms$
centred on the redshift of the galaxy around which we compute
$\rho({\bf r})$.

Figure \ref{density_fig} shows a 2D view of the VIPERS density field
(in RA and redshift) computed using the cylindrical counts-in-cells
method. Although we apply boundary corrections to the density field
computation (see Appendix \ref{app_density}), in the present work we
only use galaxies for which at least $60\%$ of the cylinder is within
the survey footprint (gaps and boundaries) presented in Fig.~\ref{fields_fig}.

With cylindrical filters we can mitigate the peculiar velocities of
galaxies in high-density regions (for example, non-linear redshift
space distortions in galaxy clusters), and by using a volume-limited
sample we measure the environment with the same tracer population at
all explored redshifts. Finally, we chose cylinders with an adaptive
radius to reach the smallest possible scales at least in high-density
regions, because it is expected that the physical processes affecting
galaxy evolution mainly occur on relatively small-scale
overdensities.

\begin{figure} \centering
\includegraphics[width=\hsize]{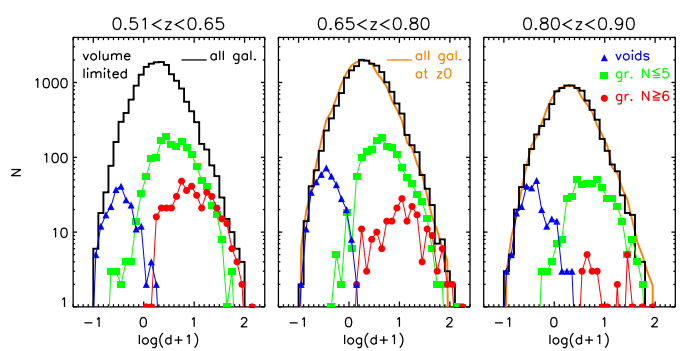}
\caption{Density distribution of the VIPERS galaxies in three redshift
  bins (three columns) for the entire sample (black histogram), for
  galaxies in voids (blue triangles), in groups with at most five members
  (green squares), and in groups with at least six members (red
  circles). See the text for the definition of groups and voids. The
  density field is computed with cylindrical filters, with a radius
  given by the fifth n.n. using the volume-limited tracers,
which are complete
  up to $z=0.9$.  To facilitate comparison, in the second and
third redshift bins the orange line is the density
  distribution of the entire sample in the first redshift bin,
  normalised to the total number of galaxies in each bin. }
\label{voids_groups_fig} 
\end{figure}

We measured the projected distance to the fifth n.n. ($D_{p,5}$),
which we used to compute the density field. For our volume-limited
tracers, $D_{p,5}$ is roughly constant with redshift. For the
volume-limited tracers limited at $M_B \leq -20.4-z$, we find
$D_{p,5}\sim5.5,~3.5,~3.2,~2.0 \mpcoh$ for
$1+\delta=0.50,~1.74,~2.10, \text{and}~5.30,$ respectively. These density values
correspond to the following key values: $1+\delta=0.50$ and 2.10 are
the median values for galaxies in voids and for all VIPERS galaxies
(see Fig.~\ref{voids_groups_fig}), while $1+\delta=1.74$ and 5.30 are
the thresholds used here to define low-density and high-density
environments (see Sect.~\ref{SM_sSFR_env}).

We verified that, as expected, $D_{p,5}$ increases for brighter
tracers, at fixed $\delta$. This is our primary motivation for
restricting our analysis to redshifts below $z=0.9$  
instead of extending it out to $z=1$ using even brighter tracers at the price of
computing the density field on much larger scales. We also verified
that although VIPERS and zCOSMOS have the same flux limit, in VIPERS
the distance to the fifth n.n. is larger than in zCOSMOS because of
its lower sampling rate and larger photometric redshift error.

As a sanity check, we verified the typical local density as measured
for galaxies in groups and in voids. Groups are identified in the flux-limited sample using a Voronoi-Delaunay-based algorithm as described
in Iovino et al. (in prep.). Here we distinguish between groups with
fewer than or at least six members. Galaxies in voids are identified as
galaxies located in the central region of spherical voids with radius
$\gtrsim10.2 \mpcoh$ and whose distance from the closest galaxy is
$\gtrsim9.8 \mpcoh$. The void-finding algorithm is the same as
presented in \cite{micheletti14}, but applied to the final VIPERS
sample. We also refer to \cite{hawken16} for a study of
voids in VIPERS.

The density distributions for all VIPERS galaxies and for galaxies in
groups and voids are shown in Fig.~\ref{voids_groups_fig} for three
redshift bins. At all redshifts, the high-density tail is mostly
populated by galaxies in groups, and the richest groups tend to reside
in the highest densities (90\% of the richest groups members fall
  in the tail of the $\sim40\%$ highest densities). In contrast, as
expected, galaxies in voids are most often found in the lowest
densities, with $90\%$ of void galaxies residing in the
  $\sim15\%$ of the lowest densities. This better agreement with void
  galaxies than with group galaxies is expected. In fact, $D_{p,5}$ in
  low densities is comparable with the typical dimension of voids,
  while in the highest densities it is still too large to be
  comparable with the small dimensions of galaxy groups and clusters
  (see above).

In Fig.~\ref{voids_groups_fig} we also observe that there is no
significant evolution of the density distribution.
\cite{kovac2010_density} showed that in the zCOSMOS bright
  sample there is also only a mild evolution of the density distribution,
  and it is mostly seen moving to $z<0.4$.

\section{Dependence of stellar mass and sSFR on the local density}\label{SM_sSFR_env}

We wish to study whether and how the stellar mass and SFR depend on
environment, and whether any dependence evolves with redshift
in the range probed by VIPERS. In particular, we focus on (a proxy of) the sSFR.

We consider the three redshift bins $0.51<z\leq0.65$, $0.65<z\leq0.8$,
and $0.8<z\leq0.9$, which were chosen because their median redshifts
are nearly equally spaced in time (with time steps of $0.6-0.7$
Gyr). In each of these bins, we consider the VIPERS sample to be
complete in stellar mass above a given mass limit $\mathcal{M}_{{\rm
    lim}}$, namely the mass limit for passive galaxies as defined in
\cite{pozzetti10}. This limit corresponds to $\log(\mathcal{M}_{{\rm
    lim}}/\mathcal{M}_\odot)=10.38$, 10.66, and 10.89 in
the three redshift bins, respectively.

We selected a sub-sample of galaxies with stellar mass above the
highest mass limit ($\log(\mathcal{M}_{{\rm
    lim}}/\mathcal{M}_\odot)=10.89$) in the entire redshift range
$0.51\leq z \leq 0.9$ . We used the first and fourth quartiles of the
density distribution of these galaxies as thresholds to define the low-density (`LD') and high-density (`HD') environments: LD
galaxies are defined by $1+\delta \leq 1.74$, and HD galaxies by
$1+\delta \geq 5.30$.  These values are very similar to those used in
\citet[][D16 from now on]{davidzon16}, where they have been derived
from an earlier smaller VIPERS data set (the first VIPERS public
release, PDR-1) and with a different SED fitting technique. This
confirms the consistency between the density field computed for the
PDR-1 and the density field computed for the final VIPERS sample (see
Appendix \ref{PDR1_vs_PDR2} for a quantitative comparison).

The average projected distance $D_{p,5}$ to the fifth n.n. for
$1+\delta = 1.74$ and $1+\delta = 5.30$ is $\sim3.5$ and $\sim2.0
\mpcoh$, respectively (see Sect.~\ref{density_field}).

\subsection{Passive and active galaxies}\label{act_pass}

We use the colour-colour diagram (NUV$-r$) vs. ($r-K$), NUV$rK$, to
define passive and star-forming (`active', from now on) galaxy
populations. In this diagram, first described in \cite{arnouts13}, the
(NUV$-r$) colour is the main tracer of recent star formation (SF); in
contrast, the ($r-K$) colour is less sensitive to SF than (NUV$-r$).
($r-K$) traces the inter-stellar medium absorption, allowing us to
separate quiescent and dusty galaxies, which show the same red
colours in a classical single-colour distribution (see
e.g. \citealp{moresco13}).  We consider a galaxy to be passive when
\begin{align}  
&  (\mathrm{NUV} - r) > 3.73 \quad  \text{and} \nonumber \\
&  (\mathrm{NUV} - r) > 1.37 \times (r - K) + 3.18 \quad  \text{and} \label{pass_eq}  \\
&  (r - K) < 1.35. \nonumber
\end{align}
These boundaries follow the definition provided by D16 (their Eq.~2),
although the values of our thresholds have been slightly modified (by
$-0.02$ and $+0.05$ mag for the (NUV$-r$) and ($r-K$) colours,
respectively) to take into account small differences in the absolute
magnitude estimates\footnote{D16 used the code {\it Hyperz}
  with a different photometric baseline. Their galaxy templates
and   the algorithm with which they computed rest-frame magnitudes are also different
  from M16b.}.

To maximise the difference between the galaxy populations, we decided
to exclude the galaxies that have intermediate colours in the NUV$rK$
plane, commonly referred to as `green valley' galaxies, from our
analysis. For this reason, our population of active galaxies is not
complementary to the passive galaxy population.  We set the upper
boundary of the active galaxies' locus to be 0.6 mag bluer in
$\mathrm{NUV} - r$ than the lower boundary of passive
galaxies. Our active population is defined as
\begin{align}  
&  (\mathrm{NUV} - r) < 3.13 \quad \text{or} \nonumber \\
&  (\mathrm{NUV} - r) < 1.37 \times (r - K) + 2.58 \quad  \text{or} \label{act_eq}  \\
&  (r - K) \geq 1.35. \nonumber
\end{align}
The condition $(r-K)>1.35$ (not used in \citealp{arnouts13})
identifies edge-on disc galaxies with a flat attenuation curve. These
are the only galaxies that can have such extreme red $(r-K)$ colours
\citep{chevallard13,moutard16a}. By using this cut in
Eqs. \ref{pass_eq} and \ref{act_eq}, we include these galaxies among
the active galaxies. We refer to \cite{fritz14} and D16 for
a more detailed discussion.

We keep the definitions in Eqs. \ref{pass_eq} and \ref{act_eq}
constant with redshift. M16b showed that these thresholds depend on
redshift, but they can be considered constant in the relatively small
redshift range $0.51<z<0.9$.

\cite{arnouts13} showed that the position of a galaxy in the NUV$rK$
plane correlates with its sSFR. This is also shown in Fig.~2 of
D16. We exploit this correlation to facilitate the comparison between our
data and the model of galaxy evolution by \cite{delucia_blaizot2007}.
The light cones we used (see Sect.~\ref{mocks}) do not have NUV,
$r,$ and K absolute magnitudes, but they do have stellar mass and SFR,
from which we can compute the sSFR. Practically, we must define some
thresholds in sSFR to define the samples of active and passive
galaxies in the model. These definitions need to correspond as closely
as possible to our classification, which is based on the NUV$rK$ diagram.

It is worthwhile to remark the following. Since the measurement of absolute
magnitudes using SED fitting techniques is more accurate than the
estimate of the SFR at the level of single galaxies, our primary
definition of the active and passive populations is the one based on
the NUV$rK$ plane. We also use the definition based on sSFR to
facilitate the comparison with the mock catalogues. It is beyond the
scope of this paper to investigate the reliability of the SFR
derived from the SED fitting in detail (however, see e.g. \citealp{conroy09}).

\begin{figure} \centering
\includegraphics[width=\hsize]{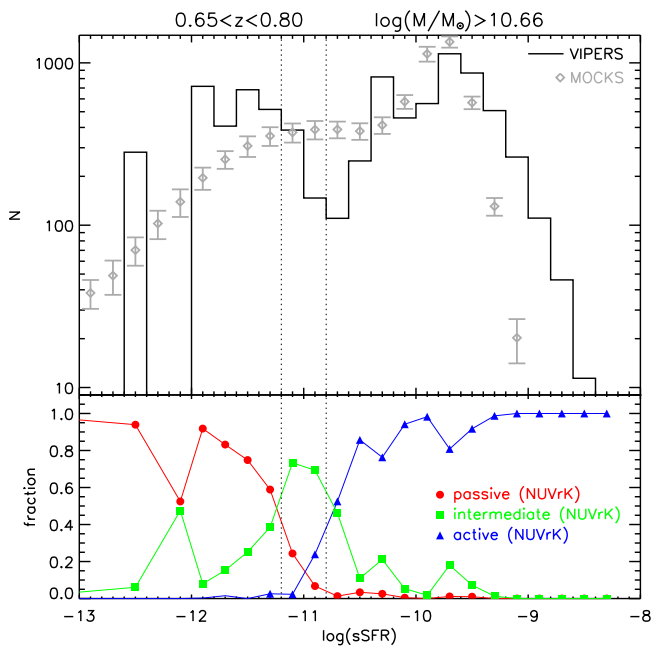}
\caption{{\it Top}. Black histogram: distribution of the sSFR ($={\rm
    SFR}/\mathcal{M}$) for the VIPERS galaxies in the redshift range
  $0.65<z<0.8$ and for $\log(\mathcal{M}_{{\rm
      lim}}/\mathcal{M}_\odot)>10.66$, which is the completeness mass
  limit in that redshift range. Diamonds: sSFR distribution in the
  \rmop\  for galaxies in the same redshift range and above the same
  mass threshold; the points with the vertical error bars correspond
  to the average and $rms$ of the 50 mocks catalogues. The sSFR
  distribution of the \rmops is normalised to have the same total
  number of galaxies as in the VIPERS sSFR distribution. Both the real
  and simulated distributions have a tail of galaxies with sSFR values
  below the lowest sSFR limit in this plot, which we do not plot for
  the sake of clarity.  {\it Bottom.}  Only for the VIPERS sample
  (same galaxies as in the top panel), fraction of passive (red
  circles) and active (blue triangles) galaxies as defined by
  Eqs.~\ref{pass_eq} and \ref{act_eq}, as a function of their sSFR. We
  also overplot the fraction of `intermediate' galaxies, i.e. those
  that do not satisfy the passive or the active definition. The
  vertical lines at $\log({\rm sSFR})=-11.2$ and $\log({\rm
    sSFR})=-10.8$ are the thresholds adopted to define passive and
  active galaxies, respectively, using the sSFR as discussed in
  Sect.~\ref{act_pass}. These results are very similar in the two
  other redshift bins (not shown). }
\label{sSFR_distrib} 
\end{figure}

In Fig.~\ref{sSFR_distrib} we show the distribution of the sSFR for
our sample in the redshift range $0.65<z<0.8$ above the mass limit
$\log(\mathcal{M}_{{\rm lim}}/\mathcal{M}_\odot)=10.66$. About
$15-20\%$ of the galaxies in this redshift range and above this mass
limit have an sSFR lower than the minimum value in the figure, and therefore
their distribution is not plotted for the sake of clarity. For any
given value of sSFR, we also show the fraction of passive and active
galaxies as defined by Eqs.~\ref{pass_eq} and \ref{act_eq}. The
correlation between this definition and the sSFR values is
evident. Moreover, we remark that this correlation is not an artefact
of the SED fitting procedure: the SFR comes directly from the
instantaneous SFR of the best-fit template, while the absolute
magnitudes are derived by minimising the template dependence by using
the observed magnitude with the closest wavelength (see
Sect.~\ref{SED}).

Figure \ref{sSFR_distrib} shows that the classification of passive and
active galaxies based on the NUV$rK$ plane roughly corresponds to
$\log({\rm sSFR})<-11.2$ and $\log({\rm sSFR})>-10.8$, respectively.
The fractions of active, intermediate, and passive galaxies as a
function of the sSFR behave in a similar way in the two other redshift
bins considered in this study, therefore we adopt the same sSFR thresholds
over the entire redshift range $0.51<z<0.9$.

In Fig.~\ref{sSFR_distrib} we also overplot the sSFR distribution in
the \rmop.  The model distribution is different from the data
distribution in several aspects. First, the tail of high sSFR is
missing in the model. Second, the valley present in the data
distribution at $\log({\rm sSFR})\sim-10.8$ appears as a plateau in
the model. It also seems to be shifted towards higher values of
sSFR. Finally, we note that in each redshift bin the tail of the sSFR
distribution below the lowest value plotted in the figure comprises
about $30\%$ of the model galaxies, but only $15-20\%$ of the VIPERS
galaxies. We refer to Appendix \ref{sSFR_model} for the
analysis of a possible cause of these discrepancies, and to
Sect.~\ref{simul} for the classification of passive and active
galaxies in the model.

\begin{figure} \centering
\includegraphics[width=7.5cm]{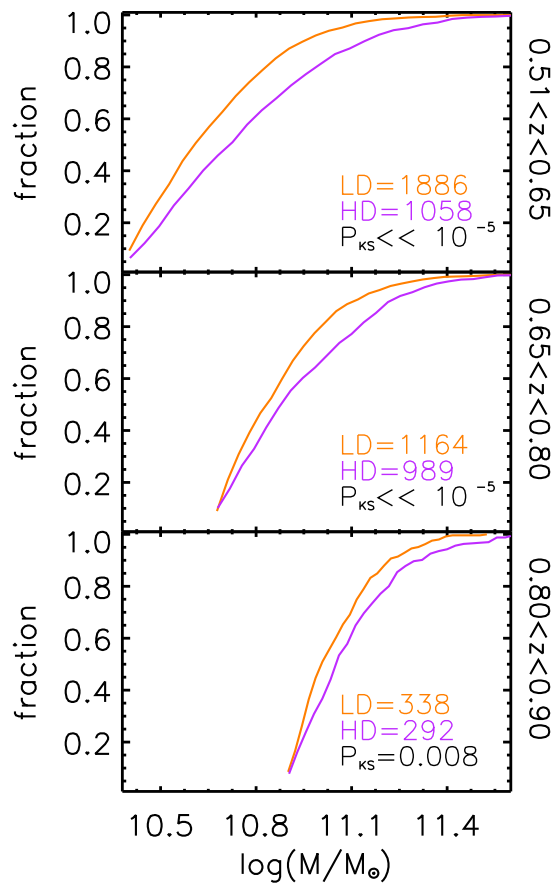}
\caption{Stellar mass cumulative distributions in three redshift bins
  ($0.51<z\leq0.65$, $0.65<z\leq0.8$, and $0.8<z\leq0.9$ from top to
  bottom), for galaxies above the mass limits $\log(\mathcal{M}_{{\rm
      lim}}/\mathcal{M}_\odot)=10.38$, 10.66, and 10.89, respectively,
  in the three redshift bins.  Orange lines show the galaxies in LD
  regions, and violet lines show galaxies in HD regions. The number of
  galaxies used in each distribution is reported in the corresponding
  panel. We use all the galaxies above the mass limit, regardless
  of their NUV$rK$ classification. In each panel we also report the
  $P_{\rm KS}$ values, i.e. the significance level in a Kolmogorov-Smirnov
  test for the null hypothesis that the LD and HD distributions are
  drawn from the same parent distribution. }
\label{mass_dens_lim} 
\end{figure}

\subsection{Stellar mass segregation in different
  environments}\label{SM_env}

It is known that galaxy stellar mass correlates with local environment
\citep[see e.g. ][]{kauffmann04,scodeggio09}. This correlation has
been extensively studied in the VIPERS survey in D16 in terms of the
galaxy stellar mass function (GSMF) in low- and high-density regions.
This dependence can also be qualitatively studied using the cumulative
distribution function of the stellar mass. Here we perform such an
analysis, primarily as a comparison with other works in the literature
that used the same tool.  As a second step, we wish to define a set
of (narrow) mass bins so that we can study how the environment affects
star formation at fixed stellar mass.

Figure~\ref{mass_dens_lim} shows the cumulative distribution function of
galaxy stellar mass in three redshift bins for galaxies above the
respective $\mathcal{M}_{{\rm lim}}$. In each redshift bin, we compare
the distributions in the two environments (LD and HD) using a
Kolmogorov-Smirnov (KS) test. In all bins we find that the
significance level $P_{\rm KS}$ for the null hypothesis, that
is, that the two
distributions are drawn from the same parent distribution, is of the
order of $\ll10^{-5}$ (with the exception of the highest redshift bin,
see below). This excludes the null hypothesis.  This is in agreement
with the different shapes of the GSMF in LD and HD regions found in
D16 (see their Fig.~4).

In more detail, at all explored redshifts the LD distribution rises
more rapidly at the lowest masses, while the HD distribution has a
more pronounced tail towards the highest stellar masses. This is in
agreement with D16, who found that the LD GSMF is steeper at low
masses and the high-mass exponential tail of the GSMF is higher in HD
regions than in LD regions. Moreover, as in D16, the LD and HD
distributions are more similar in the highest redshift bin. We
verified that the higher $P_{\rm KS}$ at $z>0.8$ is partly due to the
lower number of galaxies. Reducing the number of galaxies in the two
first redshift bins to make them equal to the third bin increases
$P_{\rm KS}$ to a few $10^{-4}$ at $z<0.8$. To verify whether the
higher $P_{\rm KS}$ at $z>0.8$ is also due to the smaller mass range
explored, we computed $P_{\rm KS}$ at $z<0.8$ imposing not only the
same number of galaxies as at $z>0.8$, but also the same stellar mass
limit $\log(\mathcal{M}_{{\rm
    lim}}/\mathcal{M}_\odot)=10.89$. Reducing the mass range, $P_{\rm
  KS}$ remains of the order of a few $10^{-4}$ at $z<0.8$. This is
still lower than $P_{\rm KS}$ at $z>0.8$. This suggests that the
dependence on environment of the high-mass tail of the stellar mass
distribution might strengthen with decreasing redshift.

\begin{table} 
  \caption{Number of active and passive galaxies in each redshift and
    stellar mass bin for the LD and HD environments. Active and
    passive galaxies are defined according to their position in the
    NUV$rK$ plane (Eqs. \ref{pass_eq} and \ref{act_eq}).}
\label{gals_nuvrk_tab} 
\centering 
\begin{tabular}{l c c } 
  \hline
  \hline   
  $\mathcal{M}$ bin [$\log(\mathcal{M}/\mathcal{M}_\odot)$]   &   act/pass (LD)  & act/pass (HD) \\
  \hline   
  \multicolumn{3}{c}{$0.51<z<0.65$} \\
  \hline
  10.38-10.66    &  666/243 & 243/111 \\
  10.66-10.89    &  285/204 & 137/132 \\
  10.89-11.09    &  102/82 &  67/96 \\ 
  11.09-11.29    &  35/41 & 24/53 \\ 
  11.29-12.00    &  4/11 & 7/39 \\
  \hline   
  \multicolumn{3}{c}{$0.65<z<0.80$} \\
  \hline
  10.66-10.89    &  365/217 & 216/199 \\
  10.89-11.09    &  143/149 &  86/122 \\ 
  11.09-11.29    &  39/54 & 51/107 \\ 
  11.29-12.00    &  4/26 & 10/48 \\
  \hline
  \multicolumn{3}{c}{$0.80<z<0.90$} \\
  \hline
  10.89-11.09    &  115/73 &  65/62 \\ 
  11.09-11.29    &  43/39 & 28/44 \\ 
  11.29-12.00    &  11/7 &  8/20 \\
  \hline
  \hline 
\end{tabular} 
\end{table}

\subsection{sSFR as a function of 
  environment}\label{sSFR_env}

In this section we investigate possible environmental effects
on the sSFR (either using the NUV$rK$ definition as a proxy, or the
SFR and stellar mass through SED fitting). In particular, we study the
ratio of the number of active to passive galaxies, $f_{\rm ap}$.  To
separate the role of stellar mass and environment, we need to study
how environment affects galaxy evolution at a fixed stellar mass. We did
this by dividing our sample into the following narrow mass bins in
$\log(\mathcal{M}/\mathcal{M}_\odot)$: 10.38-10.66, 10.66-10.89,
10.89-11.09, 11.09-11.29, and $>11.29$.

Table \ref{gals_nuvrk_tab} shows the number of active and passive
galaxies in LD and HD environments in each redshift and mass bin. In
the table, active and passive galaxies are defined according to the
NUV$rK$ diagram, but the numbers derived using the sSFR definition are
very similar.

Below we explain how we built mass-matched samples in the two
environments in each mass and redshift bin. This was to further
minimise any possible remaining difference in the stellar mass
distribution in LD and HD, even in our narrow stellar mass bins.

In each mass and redshift bin, we cut the mass distributions in the
two environments so that they have the same minimum and maximum mass
value, meaning that they cover exactly the same mass range.  Then, in each
mass and redshift bin, we (a) take the mass distribution in the
environment with the smaller number of galaxies (usually the HD
environment) as the reference mass distribution, and (b) we extract
100 samples of galaxies from the mass distribution in the other
environment, with the same mass distribution as the reference,
allowing repetitions.  Each of these 100 samples is constructed to
have the same number of galaxies as the reference.

\begin{figure*} \centering
\includegraphics[width=7.5cm]{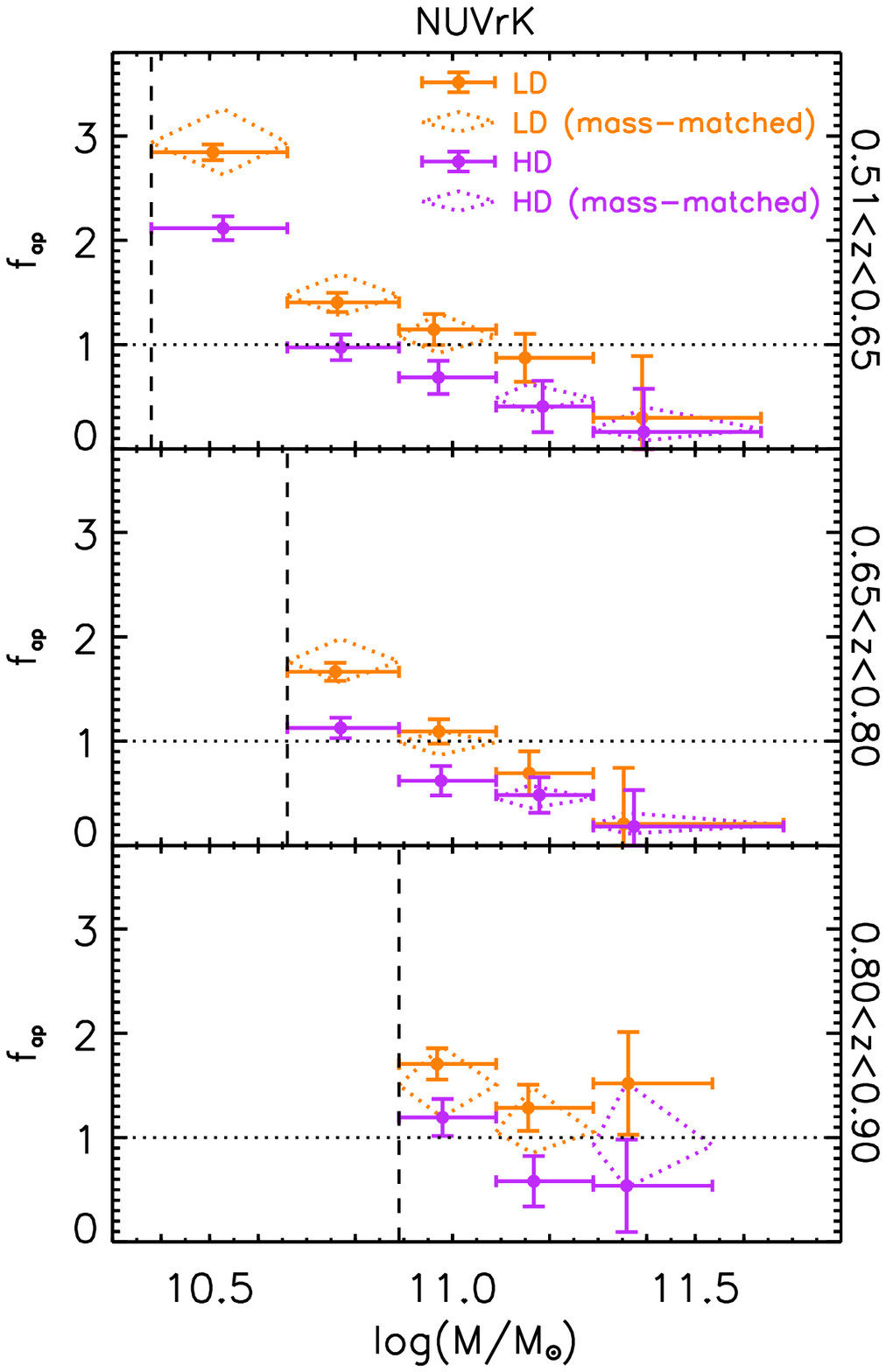}
\includegraphics[width=7.5cm]{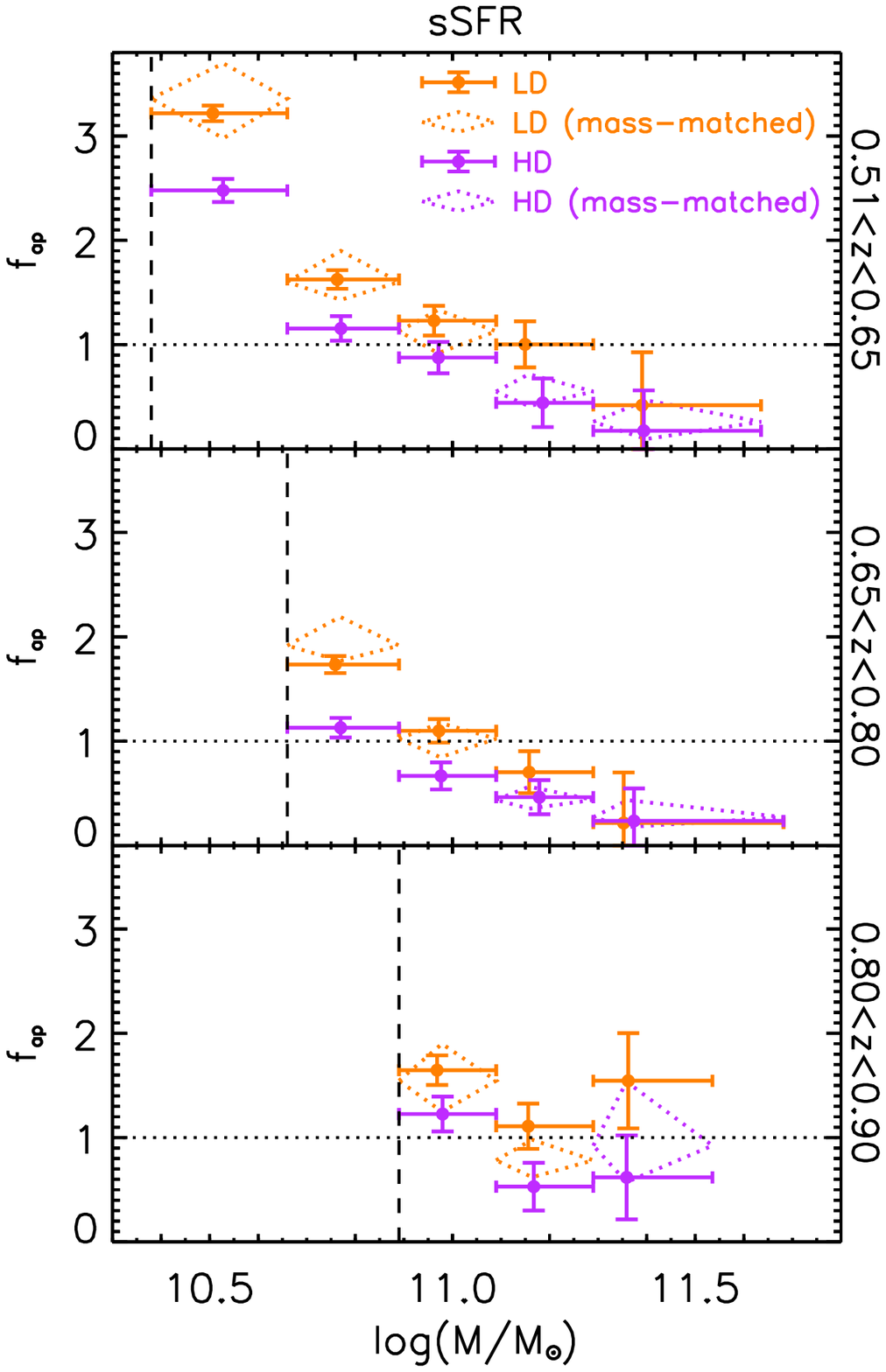}
\caption{Ratio of the number of active over passive galaxies
  ($f_{\rm ap}$) in LD (orange filled circles) and HD (violet filled
  circles) regions as a function of stellar mass in three different
  redshift bins. Active and passive galaxies are defined by means of
  the NUV$rK$ diagram {\it (left)} or according to their sSFR {\it
    (right)}. Horizontal error bars indicate the span of the mass bin,
  and the vertical error bars are derived from the propagation of the
  Poissonian noise in the counts of active and passive galaxies (if we
  use the error formula for small samples suggested in
  \citealp{gehrels86}, the error bars do not change
  significantly). The $x-$axis value is the median stellar mass of
  the sample used to compute $f_{\rm ap}$.  Dotted diamonds are for the
  mass-matched samples in the environment (LD or HD) with more
  galaxies in each mass and redshift bin. Diamonds are centred on the
  median $f_{\rm ap}$ value of the 100 mass-matched extractions, and the
  bottom and top vertices represent the 25\% and 75\% of the $f_{\rm ap}$
  distribution, respectively. The $x-$axis values is the median of the
  median stellar mass in each extraction. See text for more details.
  The vertical dashed line in each redshift bin is the corresponding
  mass limit $\mathcal{M}_{{\rm lim} }$. The dotted horizontal line at
  $f_{ap}=1$ is for reference. }
\label{ratio_dens} 
\end{figure*}

The left panel of Fig.~\ref{ratio_dens} shows $f_{\rm ap}$ as a
function of stellar mass in the redshift ranges defined above,
separating LD from HD regions. Galaxies are classified as passive and
active according to the NUV$rK$ diagram. In each mass bin, we plot
$f_{\rm ap}$ obtained by using all the galaxies in the bin (i.e.,
without applying the mass-matching method). For the environment
comprising the highest number of galaxies (LD, with the exception of
the highest mass bin) we overplot the average $f_{\rm ap}$ of the 100
mass-matched samples.

When computing $f_{\rm ap}$, in the original or mass-matched samples,
we always weight the galaxies by a factor $w$ that corresponds to the
inverse of the total sampling rate, i.e. $w=1/({\rm CSR} \times {\rm
  TSR} \times {\rm SSR})$. However, we verified that our results do
not change significantly if we do not use these weights.

We observe the following:

\begin{itemize}
 
\item[-] The median stellar mass values in each mass bin are slightly
  higher in HD than LD, suggesting that even in the narrow mass bins
  the mass distribution could be slightly different in the two
  environments. After the mass-matching, the median stellar mass
  values approach the median of the opposite environment, as
  expected.  $f_{\rm ap}$ also varies very mildly before and after the
  mass-matching.

\item[-] $f_{\rm ap}$ decreases for higher masses regardless of
  environment, as expected given the relation between stellar mass and
  SFR (e.g. \citealp{speagle14_SFR_SM} and
  \citealp{whitaker14_SFR_SM}). The only exception is the highest mass
  bin at $0.8<z<0.9$, where $f_{\rm ap}$ is similar if not higher than
  in the adjacent mass bin, although uncertainties are large.

\item[-] $f_{\rm ap}$ is higher in LD than HD environments, at all masses
  below $\log(\mathcal{M}/\mathcal{M}_\odot)=11.29$. 

\item[-] At fixed stellar mass, $f_{\rm ap}$ slightly decreases with
  redshift from $z>0.8$ to $0.65<z<0.8$ in both LD and HD, but it
  ceases to evolve at $z<0.65$.

\end{itemize}

These results are in qualitative agreement with those presented in
D16, where we separately studied the GSMF as a function of environment for active and passive galaxies. In their Fig.~5, D16 show
that the low-mass end is steeper (more negative $\alpha$) for active
galaxies. This corresponds to our $f_{\rm ap}$ increasing for low-mass
galaxies. Moreover, at $0.51<z<0.65$ the low-mass end of the passive
GSMF is much less steep in HD than in LD, which is mirrored by $f_{\rm
  ap}$ in LD increasing more steeply for lower masses.

On the other hand, we observe that in D16 the ratio of active to
passive galaxies should decrease by a factor $\sim2$ from $z\sim0.85$
to $z\sim0.55$. Based on the uncertainty on the Schechter parameters of
the GSMFs (see their Table 2), this trend is significant to a $\lesssim
2\sigma$ level.  We do not observe this redshift evolution in $f_{\rm
  ap}$, in agreement with the mild evolution of the passive and active
GSMFs in M16b.  The main reason why this trend is observed in D16 but
not M16b is that they use different SED fitting procedures, which
result in different classifications of galaxy type for a fraction of
the total sample of galaxies. Although the total GSMFs in the two
studies are in excellent agreement with each other, after dividing
their sample into active and passive galaxies, M16b found an evolution
in number density that is milder than in D16 (see Fig.~14 of M16b and
Fig.~3 of D16).  As stressed by M16b, we are currently in an era in
which large samples decrease random errors, and we can now see the
small but dominant systematics produced by different SED fitting
estimates.  In this case, the differences related to the SED fitting
procedure include different CFHTLS photometry (T005 release in D16 and
T007 in M16b), more photometric bands used in M16b, and a different
SED-fitting code.

Finally, it is worthwhile briefly discussing our findings for the highest
stellar mass bin ($\log(\mathcal{M}/\mathcal{M}_\odot)>11.29$). The
number of galaxies at such stellar masses drops steeply, and the error
on $f_{\rm ap}$ is very large. For these masses, and at $z<0.8$,
$f_{\rm ap}$ does not appear to depend on environment, but its value
is consistent with the general trend of $f_{\rm ap}$ decreasing for
higher masses, regardless of environment. At $0.8<z<0.9$, in
contrast, we do not see a clear dependence of $f_{\rm ap}$ on stellar
mass because of the smaller span in stellar mass and the relatively
high values of $f_{\rm ap}$ at
$\log(\mathcal{M}/\mathcal{M}_\odot)>11.29$ with respect to the
previous mass bins. We defer a more detailed analysis
of the properties of very massive galaxies in VIPERS to future
work.

As expected by construction, we obtain very similar results (always
within $1\sigma$) when we define active and passive galaxies using the
NUV$rK$ diagram or the sSFR thresholds, and this is crucial for the
comparison with the model.

\section{Comparison with the adopted model}\label{simul}

We make use of the 50 SAM light cones to study the dependence
of $f_{\rm ap}$ on stellar mass, redshift, and environment in the
\cite{delucia_blaizot2007} model, and we compare this to real data.

We use the \rmops and \vmops with two main aims: i) compare $f_{\rm
  ap}$ in \rmops and \vmop\  to verify that the VIPERS selection
function does not introduce any spurious signal into our measurement
of $f_{\rm ap}$, and ii) compare $f_{\rm ap}$ in the model and in the real
data to investigate which physical process(es) could be the cause of
the observed environmental trends.

We remark that these mock catalogues cover a smaller volume than the
entire VIPERS survey. Each \vmos has roughly the size of the VIPERS W4
field, which is about one-third of the whole area we probe.  For the
analysis in this section, we grouped the output (density measurement,
sSFR, stellar mass, etc.) of three \vmops at a time
for a total of 16 larger output catalogues including 48 of the
original \vmop. This simplifies the comparison with real data because
of the similar statistics in each mass and redshift bin. For
consistency, we grouped the \rmops output in the same way.

\begin{figure} \centering
\includegraphics[width=8cm]{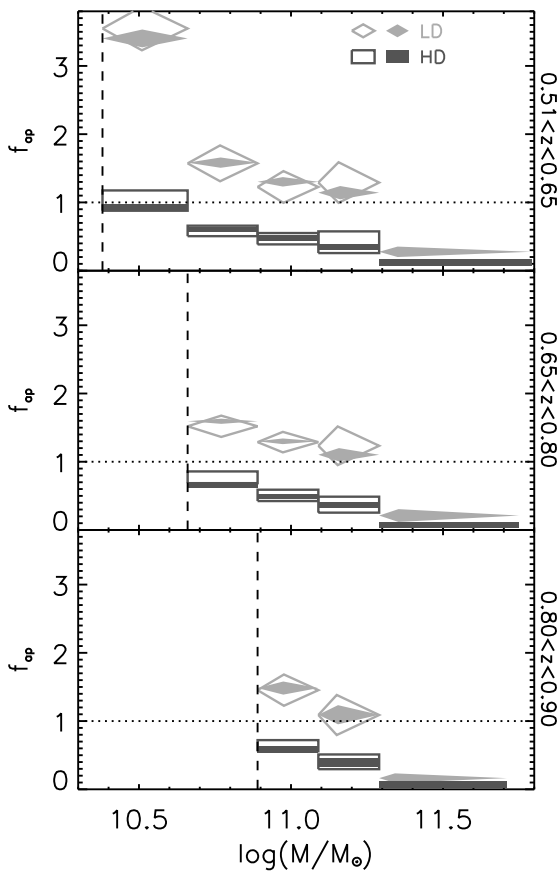}
\caption{Ratio of the number of active over passive galaxies ($f_{\rm
    ap}$) in LD (light grey) and HD (dark grey) regions in the \rmops
  (filled symbols) and in the \vmops (empty symbols) in the same
  stellar mass bins and redshift ranges as in Fig.~\ref{ratio_dens}.
  In each mass bin, for the environment with fewer galaxies (HD
    for $\log(\mathcal{M}/\mathcal{M}_\odot)<11.29$, LD otherwise) we
  plot $f_{\rm ap}$ as computed directly from the mock catalogues,
  while for the environment with more galaxies we plot $f_{\rm ap}$
  derived from the mass-matched samples. For not mass-matched
    values, symbols are centred on the $y$-axis on the mean value of
  $f_{\rm ap}$ of the 16 mock catalogues, and their height
  represents the $rms$ around the mean.  For mass-matched values,
    the $y$-axis position is computed as follows: first we compute
  the mean $f_{\rm ap}$ of the 100 mass-matched samples in each mock
  catalogue, then we average the 16 mean values. The height is given
  by the $rms$ around this mean. For all symbols, the extension on the
  $x$-axis indicates the span of the stellar mass bin. For
  $\log(\mathcal{M}/\mathcal{M}_\odot)>11.29$, we show $f_{\rm ap}$
  only for the \rmops because the statistics in the \vmops at these
  stellar masses is too low (see text).  In each panel, the vertical
  and horizontal lines are the same as in Fig.~\ref{ratio_dens}.}
\label{ratio_mill} 
\end{figure}

In the \rmops and \vmop, the local density is computed as described in
Appendix \ref{robustness_dens_app}, and LD and HD environments are
defined as for the VIPERS sample. In the mock catalogues we do not
have the absolute magnitudes in the filters needed to divide active
and passive galaxies according to their location in the NUV$rK$
plane. Instead we use their sSFR. In Fig.~\ref{sSFR_distrib} we have
shown that there is a clear difference between the distribution of the
sSFR in the \rmops and in the data. We discussed these differences in
Sect.\ref{act_pass} and Appendix \ref{sSFR_model}.  Given these
differences, we decided not to use the thresholds $\log({\rm
  sSFR})<-11.2$ and $\log({\rm sSFR})>-10.8$ (see
Sect.~\ref{act_pass}) to define passive and active galaxies in the
model, but the extremes of the sSFR distribution defined as described
in Appendix \ref{sSFR_model}.  This choice implies that $f_{ap}$ in
the model is on average in agreement with $f_{ap}$ observed in VIPERS
if we consider the entire sample regardless of environment.

We computed $f_{\rm ap}$ in the model in the same way as for the VIPERS
sample, that is to say, we built mass-matched samples in each mass and redshift
bin.  To compute $f_{\rm ap}$, the galaxies in the \vmops are weighted
by using statistical weights analogous to those used in the real
survey. The CSR is defined here as a smooth function ranging from 0 to
1 from $z=0.4$ to $z=0.6$, and it only depends on redshift. The TSR is
obtained by applying SSPOC to the mock catalogues, and the SSR is
mimicked by further downsampling the population as described in
Sect.~\ref{mocks}. In the case of the \vmop, as for the real
data, the results are not strongly dependent on the use of these
weights.

We remark that although we use the \vmops grouped 3 by 3, the
statistics of galaxies in the highest stellar mass bin is lower in
these merged mock catalogues than in the total VIPERS sample. Given
the very small numbers, we did not study this stellar mass regime in
the \vmop.

\subsection{\rmops vs \vmop}\label{rmocks_vmocks}

Figure \ref{ratio_mill} shows $f_{\rm ap}$ in the \rmops and \vmops in
the same redshift and mass bins as in the data. We verified that the mass-matched samples in the
mock catalogues are always in LD
environments, with the exception of the highest stellar masses. The
results are qualitatively similar to the results obtained with the VIPERS
data set (see Sect.~\ref{rmocks_data}), with no difference between the
\rmops and \vmop. This confirms that on average the VIPERS selection
function does not introduce any strong bias in the measurement of
$f_{\rm ap}$ as a function of environment.

The scatter around the mean values is larger in the \vmops than in the
\rmops. This could be because of the lower number of galaxies
($\sim40\%$, corresponding to the average VIPERS sampling rate), or
possibly also because of the typical uncertainties in the environment
reconstruction (see Sect.~\ref{robustness_dens_app}). We also note
that in the \vmops the dependence of $f_{\rm ap}$ on stellar mass in
LD environments almost vanishes for $\log(\mathcal{M}_{{\rm
    lim}}/\mathcal{M}_\odot)>10.66$, while it is mild but evident in
the \rmop. Finally, we do not find any dependence of $f_{\rm ap}$ on
redshift at fixed stellar mass in either the \vmops or in the
\rmop.

Given the differences between \rmops and \vmop, we expect that the
trends of $f_{\rm ap}$ with stellar mass observed in the VIPERS sample
in Fig.~\ref{ratio_dens} are weaker than the true trends (especially
in LD). We also expect that the difference of $f_{\rm ap}$ in LD and
HD is less significant because of both the low(er) statistics and the
errors in the environment reconstruction.  Moreover, the lack of
dependence of $f_{\rm ap}$ on redshift at fixed stellar mass
that we found
in the VIPERS data set does not seem to be due to the VIPERS
selection function.

From now on, since we have verified that the $f_{\rm ap}$ behaviour is
compatible in the \rmops and \vmop, we only use
the \rmop\ for simplicity.

\begin{figure} \centering
\includegraphics[width=8cm]{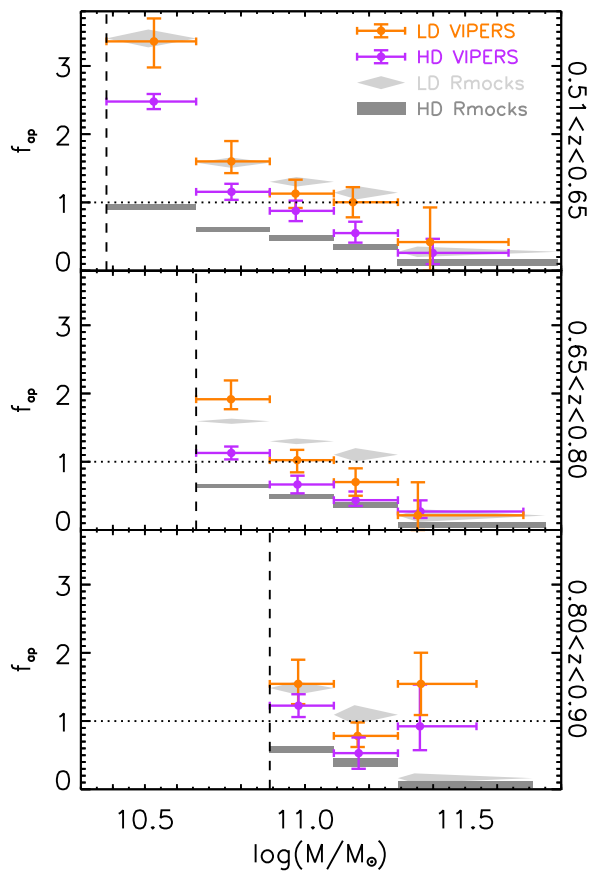}
\caption{Ratio of the number of active over passive galaxies ($f_{\rm
    ap}$) in VIPERS real data (orange and violet symbols) and in the
  \rmops (filled rectangles and diamonds). The points for the
    \rmops are the same as in Fig.~\ref{ratio_mill}. The points for the VIPERS
    sample are taken from Fig.~\ref{ratio_dens}, but for the sake of
    simplicity, in each mass bin for the environment with more
    galaxies (LD for $\log(\mathcal{M}/\mathcal{M}_\odot)<11.29$, HD
    otherwise) we plot only $f_{\rm ap}$ as derived from the
    mass-matched samples, instead of also plotting the original value
    as in Fig.~\ref{ratio_dens}.} 
\label{ratio_real_mill} 
\end{figure}

\subsection{Comparison between data and \rmops}\label{rmocks_data}

Figure \ref{ratio_real_mill} shows the VIPERS $f_{\rm ap}$ with
$f_{\rm ap}$ of the \rmops overplotted.  The trends are qualitatively
similar, with $f_{\rm ap}$ decreasing for higher stellar masses in
both LD and HD, and with $f_{\rm ap}$ higher in LD than in HD at all
masses at $\log(\mathcal{M}/\mathcal{M}_\odot)<11.29$. The peculiarity
of $f_{\rm ap}$ for $\log(\mathcal{M}/\mathcal{M}_\odot)>11.29$ at
$z>0.8$ in the data, where it does not follow the trend with stellar
mass, is underlined by the fact that in the model $f_{\rm ap}$ is
lower that in the previous mass bins.

The most remarkable result shown in Fig.~\ref{ratio_real_mill} is that
the model $f_{\rm ap}$ in LD environments is very similar to the one
in the data, while $f_{\rm ap}$ in HD environments underestimates the
observed $f_{\rm ap}$ for $\log(\mathcal{M}/\mathcal{M}_\odot)<11.09$,
and this underestimate becomes more severe as mass decreases.

\section{Discussion}\label{discussion}

Unfortunately, it is not straightforward to compare different works in
the field of environmental effects on galaxy evolution. Several galaxy
classification systems have been adopted (by colour, SF, morphology,
etc.), and the definition of environment itself can vary from analysis
to analysis.  Nevertheless, our results can be put into perspective by
discussing the implications of our findings in the framework defined
by the literature (Sect.~\ref{discussion_data}). Moreover, our
analysis with mock galaxy catalogues allows us to gain insights into
the physical processes responsible for the disagreement found between
our observational measurements and predictions from galaxy formation
models (Sect.~\ref{discussion_models}).

\subsection{VIPERS in context}
\label{discussion_data}
 
The main results of the present work are the clear preference for
massive galaxies to reside in the densest environments
(Fig.~\ref{mass_dens_lim}) and the fact that even at fixed stellar
mass, these overdensities host a lower percentage of active galaxies
than in low-density regions (Fig.~\ref{ratio_dens}).  Although at
different confidence levels, this second result seems to hold for all
the stellar masses analysed in this paper up to
$\log(\mathcal{M}/\mathcal{M}_\odot)=11.29$. Above this limit we do
not detect any significant dependence upon environment. We remark that
these results are in qualitative agreement with the work by
\cite{malavasi17}, where we reconstruct the VIPERS large-scale
filamentary structure and find that the most massive (or quiescent)
galaxies are closer to filaments than less massive (or active) ones.

The stellar mass segregation that we find is expected to be a direct
consequence of structure formation and hierarchical halo assembly
\citep[see][]{mo_white97}.  Although some previous works, using
different approaches, found the same environmental trend
\citep[e.g.][]{abbas05,scodeggio2009_VIMOS,wetzel12,vanderBurg14}, one
recent analysis \citep{kafle16} challenges this scenario by finding no
evidence of stellar mass segregation in SDSS galaxy groups.  Such
contrasting conclusions could be explained by differences between
group-finding algorithms \citep[see][]{campbell15} and/or input
datasets (e.g., different proxy to recover halo masses) used by
different authors.  These findings again highlight the effects that
different environment definitions have on the analysis of
environmental effects on galaxy evolution. The debate surrounding the
segregation of massive galaxies is still very much active.

In this respect, we find (Fig.~\ref{ratio_real_mill}) that at
$\log(\mathcal{M/M}_\odot)\gtrsim11.3$ the SAM mocks correctly
reproduce our survey, at least up to $z=0.8$.  According to the model,
most of these massive galaxies are the central object in their halo,
both in LD regions ($>95\%$) and HD ($>80\%$).  Despite the mass
segregation effect discussed above, their number in the LD regions is
not negligible (numbers similar to those of the VIPERS sample, see
Table \ref{gals_nuvrk_tab}). Whilst we defer a detailed study of such
massive objects to a future paper in this series, here we would like
to emphasise the dependence of stellar-to-halo mass ratio on the
large-scale environment. This is often neglected in halo occupation
distribution models \citep[as pointed out in][]{tonnesen15}. We remark
that this topic, referred to as `assembly bias', is hotly debated in
the literature, as shown by its long history.  For instance,
\cite{lemson99} were among the first who attempted to detect it
(although they did not find it because this effect is more important
for lower mass systems, which were not well resolved in earlier
simulations), and it was only first measured by \cite{gao05} using
N-body simulations.

Concerning the relative contribution of active vs passive galaxies, we
find that their ratio $f_\mathrm{ap}$ is lower in denser
environments, in agreement with what has been observed in the local
Universe \citep[e.g][]{baldry06} and at intermediate redshifts
\citep[e.g.~DEEP2, VVDS, and zCOSMOS, see
][respectively]{cooper08_SFH,cucciati2006,cucciati10}.  These authors
all used a method similar to ours, based on the $n$-th nearest
neighbour or apertures with fixed radius, to define the local
environment on comoving scales $\leqslant 5$ $h^{-1}$Mpc.
\citet{burton13} find the same results in GAMA
\citep[$0.02<z<0.5$,][]{baldry10} by using Voronoi tessellation and
far-infrared emission as a proxy of SFR.  A similar dependence of
galaxy populations on environment is observed in galaxy clusters,
where the fraction of active galaxies increases as a function of
cluster-centric radius \citep[e.g.][]{treu03,raichoor14,haines15}.  In
agreement with our analyses, \citet{patel11} show that this
environmental trend does not vary when different
SFR estimators are used, and
the decline of the SFR in groups and clusters at $0.6<z<0.9$ is indeed
caused by a smaller fraction of active galaxies and not by a change
in the global sSFR distribution.
 
\begin{figure*} \centering
\includegraphics[width=6cm]{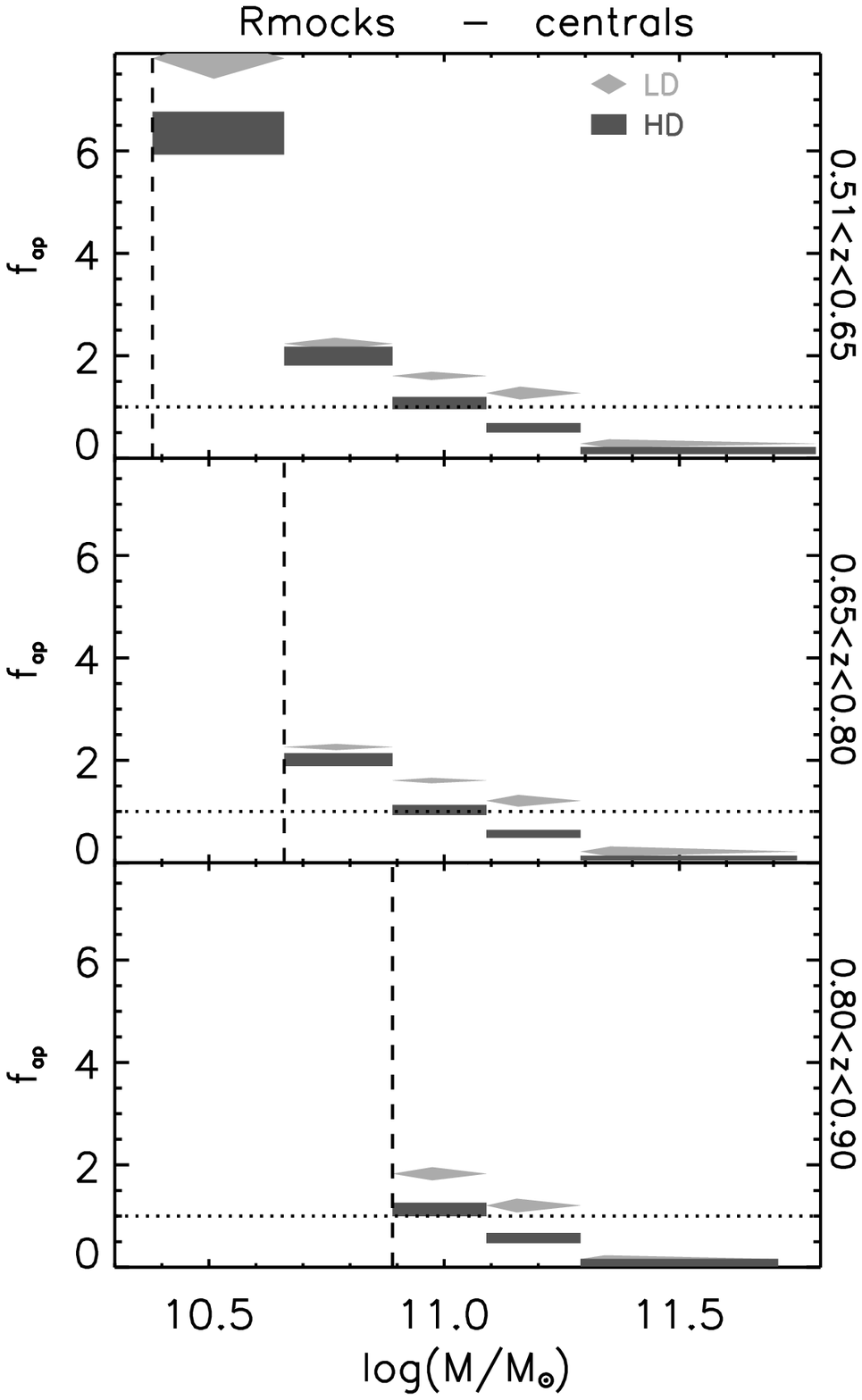}
\includegraphics[width=6cm]{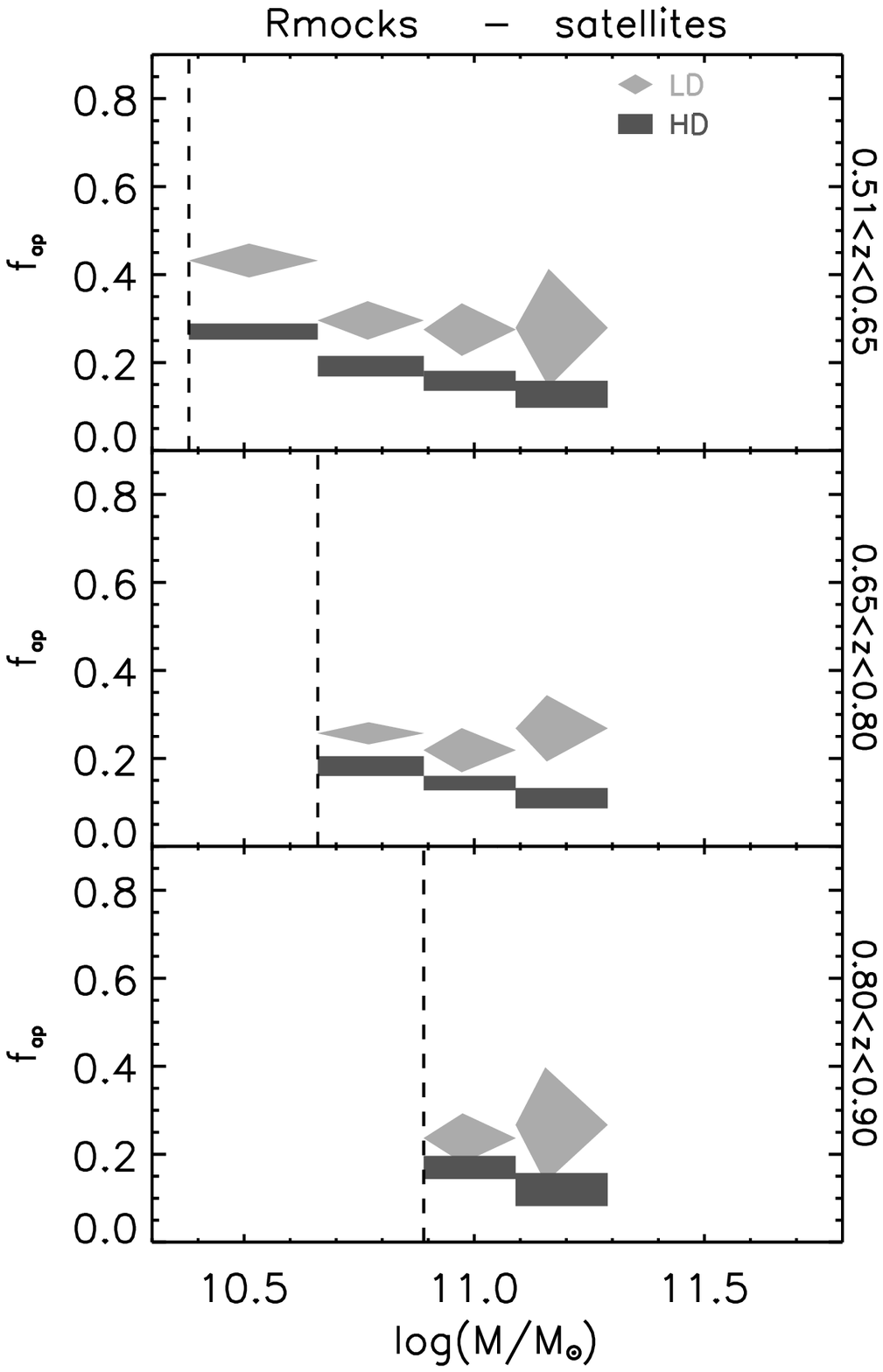}
\includegraphics[width=6cm]{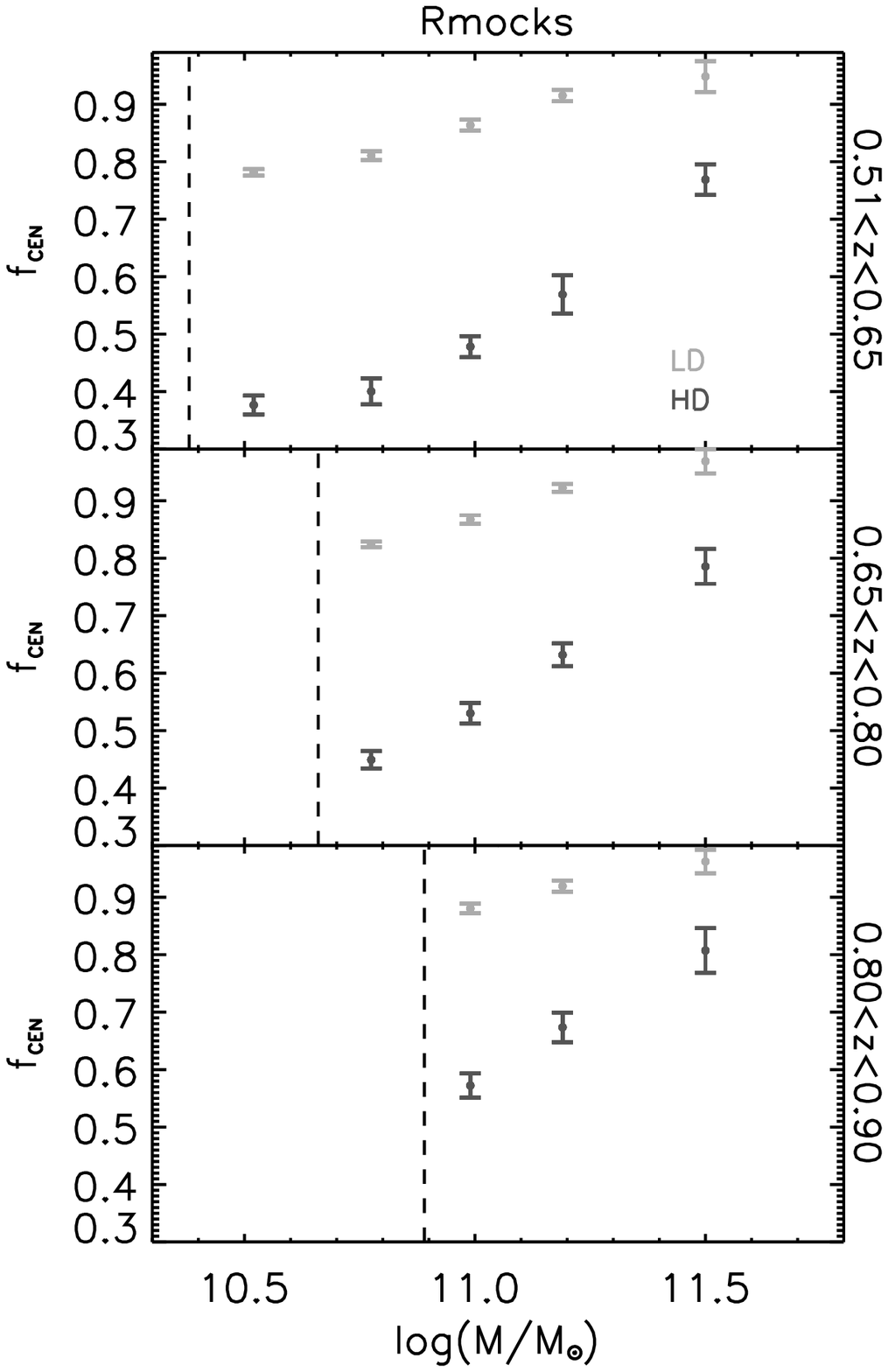}
\caption{{\it Left.} Fraction $f_{\rm ap}$ in the \rmops (similar to the filled
  symbols in Fig.~\ref{ratio_mill}), but considering only central
  galaxies. {\it Middle.} As in the {\it left} panel, but considering
  only satellite galaxies. Note the different $y$-axis ranges in the
  {\it left} and {\it middle} panels. We plot $f_{\rm ap}$ only in the
  mass bins where there are at least 20 galaxies per light cone. {\it
    Right.} Fraction of central galaxies in the \rmop in the same
  redshift ranges and stellar mass bins as in the {\it left} and {\it
    middle} panels. The fraction is the mean value of the 16 \rmops
  and the vertical error bar is the rms around the mean. The fraction
  refers to the total number of galaxies, regardless of whether they are
  classified passive, intermediate, or active }
\label{ratio_mill_cen_sat} 
\end{figure*}

The decrease of $f_\mathrm{ap}$ as mass increases is (qualitatively)
consistent with the increase of the `red sequence fraction' adopted by
\citet{baldry06} to analyse SDSS data.  From SDSS group
classification, \citet{wetzel12} show that such a dependence on
stellar mass vanishes when only satellite galaxies are considered
\citep[although see also][]{delucia12}. Similarly, as we show in
Sect. \ref{discussion_models}, in our mock catalogues we find that
satellites galaxies show a less evident dependence of $f_{\rm ap}$ on
stellar mass than central galaxies. We have not yet attempted to
identify satellites in the VIPERS sample, but if we assume that their
fraction is a function of stellar mass, as our mock catalogues indicate
(Fig.~\ref{ratio_mill_cen_sat}, right panel), then the observed
$f_\mathrm{ap}$ should become less dependent on $\mathcal{M}$.

Several scenarios have been proposed to explain the environmental
trends we observe.  Quenching mechanisms like ram pressure stripping
or galaxy-galaxy interactions \citep[for an exhaustive review,
see][]{boselli_gavazzi06} are more likely to occur in overdense
regions (each process in a typical environment and on typical
timescales), while other processes (e.g., AGN feedback) seem to be
independent of environment \citep[at least in the VIPERS redshift
range, see][]{rumbaugh17}.  Moreover, the latter should be dominant at
high masses \citep[e.g.][]{peng10,gabor14} where indeed we find
similar $f_\mathrm{ap}$ fractions in the two opposite environments.
   
In this respect, a missing piece of the mosaic is a precise
determination of the quenching timescale, which is still debated
\citep[see e.g.~the discussion in][]{haines15,moutard16b}.  We address
this point below, trying to reconcile the results of our mock
catalogues with VIPERS. Our analysis spans over an epoch ($0.5<z<0.9$)
when the star formation rate density (SFRD) of the Universe has
already begun to decline \citep[see ][]{cucciati12,madauDickinson14},
and we do see an environmental signature in galaxy quenching across
the entire redshift range explored. On one hand, it is of paramount
importance to understand how much environment-driven processes
contribute to the SFRD decline. On the other hand, the debate is still
open on the possible link between the peak of the SFRD at $z\sim1.5-2$
and the absence, or even inversion, of the correlation between star
formation activity and local density at such redshift
\citep[e.g.][]{mortlock15,alberts16,
  cucciati2006,elbaz07,cooper08_SFH, grutzbauch11a}.

\subsection{Insights from the model}\label{discussion_models}

The remarkable agreement of $f_{\rm ap}$ in LD in the model and in the
data and the under-prediction of $f_{\rm ap}$ in the model in HD
suggest that a class of galaxies (e.g. the satellite galaxies, which
reside mainly in HD regions, see below) with erroneous properties
exists in the models, or a class with the correct properties, but with
incorrect number counts.

We already know that the model by \cite{delucia_blaizot2007},
similarly to most models available today, overpredicts the number of
low-mass passive galaxies
\citep[e.g. ][]{wang07,fontanot09,weinmann10,henriques13,cucciati12}. Here
we show that this excess of passive galaxies is more important in
high-density regions. This was expected, to some extent, because such
low-mass passive galaxies are preferentially satellites (also see
below), which reside in the high-density regions given by galaxy
groups and clusters. In the models, galaxies undergo some
physically motivated processes when they become satellites. These have
the net effect of quenching the SF. It has been pointed out in
the literature that these processes are probably modelled to be too
strong or too quick, causing an overprediction of relatively low-mass
passive galaxies. Some more recent models have tried to
mitigate this quenching \citep[e.g. ][]{font08,guo10}, and also
obtained improvements on the environmental effects on galaxy star
  formation \citep[e.g. ][]{henriques16}. However, the excess of
low-mass passive galaxies might not just be a problem resulting from
over-efficient hot-gas stripping when a galaxy becomes a
satellite. For instance, \cite{hirschmann16_GAEA} show that this
excess can be significantly reduced by modifying the stellar feedback
scheme (i.e. in this case, the properties of galaxies change at the
time they become satellites), and argue that the quenching timescales
are not primarily determined by environmental processes.

We can verify whether the low $f_{\rm ap}$ values in HD in the \rmops
are mainly due to the satellite galaxies by separating them from the
central galaxies. In Fig.~\ref{ratio_mill_cen_sat} we show $f_{\rm
  ap}$ in the \rmops using only central (left panel) or only satellite
galaxies (middle panel). We find the following:

\begin{itemize}

\item[-] $f_{\rm ap}$ is much higher for central galaxies than for
  satellites for all stellar masses, environments, and redshift
  ranges. This is especially true for
  $\log(\mathcal{M}/\mathcal{M}_\odot)<11.09$.

\item[-] $f_{\rm ap}$ is higher in LD than HD environments, at all masses
  below $\log(\mathcal{M}/\mathcal{M}_\odot)=11.29$ for both centrals
  and satellites.

\item[-] $f_{\rm ap}$ in HD increases for lower stellar masses for
  both central and satellite galaxies, but this increase is much
  steeper for centrals than for satellites.

\item[-] $f_{\rm ap}$ in LD shows the same increase as in HD at
  lower stellar masses for central galaxies, while for satellite
  galaxies it seems to depend very weakly (if at all) on stellar mass,
  at least for $\log(\mathcal{M}/\mathcal{M}_\odot)>10.66$.

\item[-] In HD, $f_{\rm ap}$ is always higher for central galaxies
  than for the total sample (see Fig.~\ref{ratio_mill}), while for
  satellite galaxies it is always lower than for the total
  sample. Moreover, in HD the increase of $f_{\rm ap}$ for the central
  galaxies towards lower masses is much steeper than that of the total
  sample.
 
\end{itemize}

\begin{figure} \centering
\includegraphics[width=8cm]{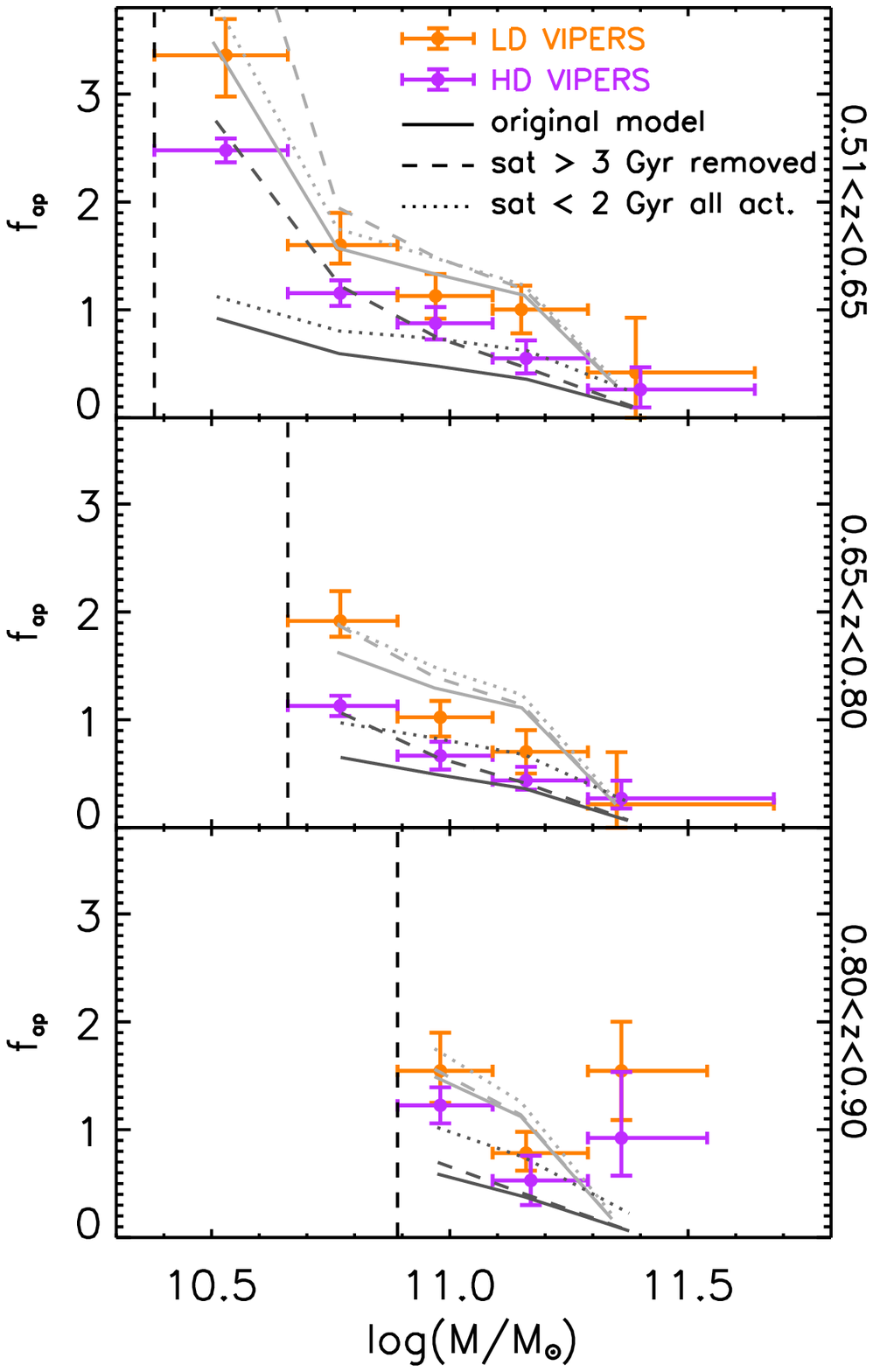}
\caption{Fraction $f_{\rm ap}$ in the VIPERS data (orange and violet
  crosses) and in the model (grey lines). Crosses are the same as in
  Fig.~\ref{ratio_real_mill}.  Solid grey lines refer to \rmops and
  correspond to the grey polygons of Fig.~\ref{ratio_mill}. Dashed
  lines refer to \rmops, but are computed after removing satellite
  galaxies that became satellites more than 3 Gyr before. The dotted
  lines refer to \rmops, but are computed considering all the `young'
  satellite galaxies (i.e., that became satellites less than 2 Gyr
  before) to be active. See the text for more details.}
\label{ratio_mill_tsat} 
\end{figure}

The low $f_{\rm ap}$ in HD in the total sample at the lowest masses is
clearly due to the satellite galaxies. The SF is much more quenched in
satellite than in the central galaxies, and they are more abundant in
HD at these masses (see the rightmost panel of
Fig.~\ref{ratio_mill_cen_sat}).  In contrast, in LD environments the
total sample is dominated by central galaxies.

The very low fraction of active galaxies in HD in the model with
respect to the real data could be due to an excessive quenching of
satellite galaxies or to an excessive number of satellite
galaxies. In contrast, it seems that in LD the model is able to
reproduce the correct balance of satellite and central, and their
respective quenching, in order to agree with the data.  This result is
helpful for improving the inclusion of environment-driven processes in
the model. However, for a more meaningful comparison it would be
necessary to identify central and satellite galaxies in the real data
as well, which is not an easy task (although see e.g. \citealp{kovac14}
for a central-satellite classification at $z\sim0.7$); we defer this
analysis of the VIPERS sample to future work.

It is not the aim of this work to perform a detailed analysis of the
environmental history of galaxies in the model. However, we performed
a simple test to verify two hypotheses. We assumed that there
are too many low-mass passive satellites in the simulation because
either i) they survived (when they should not have) the disruption of
multiple mergers or encounters, or ii) they are already passive (when they
should not be) because the quenching mechanisms are too
fast or strong. In the first case we investigate the total number of
satellites in the model, in the second case we investigate their star
formation activity.

We address these two possibilities in the following way. We study the
merger tree of the satellite galaxies, and we find the time $t_{\rm
  sat}$ at which they became satellites. The time $t_{\rm sat}$ is
defined as the last time when the main progenitor of the satellite
galaxy is a central galaxy. As a first check, we verified that at all
stellar masses and redshifts explored, galaxies in HD became
satellites on average 0.5-1 Gyr before galaxies in LD
environments. Secondly, we recomputed $f_{\rm ap}$ in the \rmops by
modifying the sample of satellite galaxies in two ways: i) we removed
all the galaxies that became satellites more than $X$ Gyr before the
time of observation from the sample, to take into account the fact
that some of them should have been destroyed); ii) we considered all
the passive satellites that became satellites less than $Y$ Gyr
before to be active, to compensate for the possibility that the quenching
they have undergone at the moment of becoming satellites was too
strong or quick.

We explored different values of $X$ and $Y$. In
Fig.~\ref{ratio_mill_tsat} we show $f_{\rm ap}$ in the \rmops for
$X=3$ Gyr and $Y=2$ Gyr, that is to say, the values for which in HD the model
$f_{\rm ap}$ is more similar to the data $f_{\rm ap}$. We remark that
to keep this toy model as simple as possible, we used the same $X$ and
$Y$ at all stellar masses and at all redshifts, and in both
environments.

In the first case (the removal of satellites), the model $f_{\rm ap}$
in HD is in agreement with observations when we exclude galaxies that
became satellites more than 3 Gyr before from the sample. This is true
at all masses in the redshift range $0.51<z<0.8$, but at $z>0.8$ we
would need to remove even more `recent' satellites ($X<3$ Gyr) to make
the model $f_{\rm ap}$ agree with the observed $f_{\rm ap}$ (we would
need to remove all the galaxies that became satellites more than 1 Gyr
before).  Figure \ref{ratio_mill_tsat} also shows that by removing
such satellites regardless of their environment, we also increase
$f_{\rm ap}$ in LD (especially at the lowest redshift), which now
would over-predict the observed one. We note that the fraction of
removed satellites is non-negligible and strongly depends on stellar
mass. In the lowest redshift bin, in HD it ranges from $\sim50\%$ to
$\sim30\%$ going from the lowest to the highest stellar masses. These
fractions become smaller and smaller at higher redshift. These trends,
together with the fact that the satellite fraction is higher at lower
masses, explain why $f_{\rm ap}$ changes much more at low masses and
at low redshift.  Moreover, they are smaller in LD at all redshifts,
but still non-negligible at $0.51<z<0.65$, where we remove $\sim 40\%$
of the satellites with the lowest masses, so that $f_{\rm ap}$ also
increases in LD in this redshift and mass regime.

In contrast, by moving satellites from the passive to the active
population, for $Y=2$ Gyr the model $f_{\rm ap}$ in HD is in better
agreement with the observed one at $z>0.65$, while at $z<0.65$ $f_{\rm
  ap}$ in HD is still too low for
$\log(\mathcal{M}/\mathcal{M}_\odot)<10.89$. It is possible to
increase $f_{\rm ap}$ at low masses by increasing $Y$, but in this way
$f_{\rm ap}$ in HD would become too high at higher stellar masses, and
at all masses at higher redshift (we refer e.g. to
\citealp{wetzel12} and \citealp{hirschmann14} for a discussion of the
mass dependence of the quenching timescales). For $Y=2$ Gyr, the
fraction of passive satellites that we move into the star-forming
population increases with increasing stellar mass and (mildly) with
increasing redshift. Moreover, it is similar in HD and LD.  For
instance, it ranges from $\sim15\%$ to $\sim45\%$ from the lowest to
the highest stellar masses in the lowest redshift bin. Given that we
move a larger satellite fraction at higher masses, but the overall
satellite contribution (right panel of Fig.~\ref{ratio_mill_cen_sat})
increases for lower masses, the net effect is that $f_{\rm ap}$
increases roughly homogeneously at all stellar masses.

In conclusion, both these changes to the model significantly increase
the agreement between the model and the data, even if neither of them
is able to reproduce the data over the entire range of stellar mass,
environment, and redshift.  Clearly, by considering the two cases
separately, we make an over-simplistic assumption. It is probable that
the problem is (at the very least) two-fold: the model has too many low-mass satellites, and too many of them (even of
the `right' ones) have been too heavily quenched. Moreover, these
effects are likely to depend in different ways on redshift, stellar
mass, and environment. In summary, this simple test tells us that in
order to fine-tune a model of galaxy formation, we need to provide
solid observational constraints considering the local environment of
galaxies, possibly in a broad redshift range to better model the
quenching timescales, which is not an easy task.

\section{Summary and conclusions}\label{summary}

VIPERS is a flux-limited survey ($i<22.5$) conceived to be an analogue
of the local 2dFDRS, but at higher redshift ($0.5<z<1.2$).  Thanks to
the large volume explored, we can study galaxy evolution with accurate
statistics and also perform detailed analyses of rare galaxy
populations, such as the most massive galaxies. In this work, we used
the final VIPERS release to study how environment affects the sSFR in
galaxies. We defined the environment as the galaxy density contrast
computed using cylindrical top-hat filters, on scales corresponding to
the fifth nearest neighbour, by using spectroscopic redshifts and
additionally photometric redshifts for the galaxies without reliable
spectroscopic measurements. We verified the reliability of the density
field reconstruction using galaxy mock catalogues mimicking the VIPERS
observational strategy. The main results of our analysis are as
follows:

\begin{itemize}

\item[-] More massive galaxies tend to reside in high-density regions
  (HD) rather than in low-density regions (LD). This is true throughout the
  entire explored redshift range ($0.51<z<0.9$), in agreement with
  previous works.
\item[-] We defined active and passive galaxies by means of the
  (NUV$-r$) vs ($r-K$) colour-colour diagram, and we studied how the
  fraction $f_{\rm ap}$ of active over passive galaxies depends on
  environment at fixed stellar mass. We found that $f_{\rm ap}$ is
  higher in LD than in HD, from the lowest stellar masses explored
  ($\log(\mathcal{M}/\mathcal{M}_\odot)=10.38$) at least up to
  $\log(\mathcal{M}/\mathcal{M}_\odot)\sim11.3$, although with
  decreasing significance from lower to higher masses.
\item[-] We performed the same analysis on VIPERS-like mock galaxy
  catalogues, based on the model of galaxy formation and evolution by
  \cite{delucia_blaizot2007}. We found that the model $f_{\rm ap}$
  reproduces the observed $f_{\rm ap}$ in LD VIPERS environments well,
  but it underpredicts $f_{\rm ap}$ in HD environments. This is
  especially the case for the lowest stellar masses. We verified that
  this underprediction is mainly due to an excess of low-mass passive
  satellite galaxies in the model.
\item[-] By studying the time $t_{\rm sat}$ when galaxies became
  satellites, we verified that $f_{\rm ap}$ in the model could become
  too low in two different ways: (i) through the spurious presence of
  `old' satellites that are still present in the simulation more than 3 Gyr
  after $t_{\rm sat}$, which should have been disrupted by close
  encounters; (ii) by too rapid or strong quenching processes in `young'
  satellites that terminate star formation in galaxies that have
  recently ($<2$ Gyr) become satellites.
\end{itemize}

Our results are in very good agreement with other works performed with
the VIPERS sample, that is, with the study of the galaxy stellar mass
function per environment \citep[][based on a previous VIPERS
release]{davidzon16} and the study of galaxy properties with relation
to distance from filaments describing the VIPERS large-scale structure
\citep[][based on the same VIPERS release as this work, but with a
different definition of environment]{malavasi17}.

It is difficult to perform a quantitative comparison of our results
with similar works in the literature because of the different
environment definitions (or the same definition, but different tracer
samples) and the different galaxy-type classification. Nevertheless,
this is the first time that environmental effects on high-mass
galaxies are clearly seen up to $z\sim0.9$. This result throws
doubts on the hypothesis that stellar mass is the only driver of SF quenching in
galaxies with $\log(\mathcal{M}/\mathcal{M}_\odot)\gtrsim10.6$
\citep[see e.g.][]{peng10}.

Our study extends up to $z\sim0.9$, and we find an environmental
signature in galaxy quenching across the entire redshift range
(although less clear at $0.8<z<0.9$ because of the lower mass range
probed). This is an epoch at which the star formation rate density (SFRD)
of the Universe has begun to decline. It is of paramount
importance to understand the reasons for this decline and to which
extent environment-driven processes contribute to it. In this respect,
solid environmental studies at $z\lesssim1-1.5$ could give important
clues to define this picture. Needless to say, we also need a
continuous feedback between observations and models of galaxy
formation and evolution. Our comparison with the model proposed by De
Lucia \& Blaizot and our simple toy model based on the time when
galaxies became satellites suggests that we need to provide solid
observational constraints also based on galaxy local environment,
possibly in a broad redshift range to better model quenching
timescales. We need a combined effort to develop in parallel more
efficient instruments and surveys to study the local galaxy environment at
high redshift (for an example of what can be done in this regard with
Euclid, see e.g.  \citealp{cucciati16a}) and better fine-tuned models
to understand how and at which pace galaxies evolve through the cosmic
large-scale structure of the Universe.

%****************************************************************************

\begin{acknowledgements} We acknowledge the crucial contribution of
  the ESO staff for the management of service observations. In
  particular, we are deeply grateful to M. Hilker for his constant
  help and support of this program. We thank the anonymous referee for
  the useful comments and suggestions. Italian participation to VIPERS
  has been funded by INAF through PRIN 2008, 2010, and 2014
  programs. LG, AJH, and BRG acknowledge support from the European
  Research Council through grant n.~291521. OLF acknowledges support
  from the European Research Council through grant n.~268107.  JAP
  acknowledges support of the European Research Council through grant
  n.~67093. RT acknowledges financial support from the European
  Research Council through grant n.~202686. AP, KM, and JK have been
  supported by the National Science Centre (grants
  UMO-2012/07/B/ST9/04425 and UMO-2013/09/D/ST9/04030). WJP is also
  grateful for support from the UK Science and Technology Facilities
  Council through the grant ST/I001204/1. EB, FM and LM acknowledge
  the support from grants ASI-INAF I/023/12/0 and PRIN MIUR
  2010-2011. LM also acknowledges financial support from PRIN INAF
  2012. SDLT and MP acknowledge the support of the OCEVU Labex
  (ANR-11-LABX-0060) and the A*MIDEX project (ANR-11-IDEX-0001-02)
  funded by the "Investissements d'Avenir" French government program
  managed by the ANR and the Programme National Galaxies et Cosmologie
  (PNCG). TM and SA acknowledge financial support from the ANR Spin(e)
  through the french grant ANR-13-BS05-0005. Research conducted within
  the scope of the HECOLS International Associated Laboratory,
  supported in part by the Polish NCN grant DEC-2013/08/M/ST9/00664.
\end{acknowledgements}

%\end{thebibliography}
\bibliographystyle{aa}
\bibliography{biblio,biblio_VIPERS_v2}

\appendix

\section{Density field reconstruction}\label{app_density}

We summarise the procedure used to derive the VIPERS density field and
to assess the reliability of its computation.

\subsection{Local density $\rho$}\label{app_rho}

Several factors need to be taken into account when computing the local
density $\rho({\bf r})$ (Eq.~\ref{rho_eq}) and its mean value $\langle
\rho({\bf r}(z)) \rangle$. A detailed description of the procedure and
its parameters are described in \cite{kovac2010_density}. Here we
discuss the most relevant issues that we considered for our
analysis.

{\bf 1) Tracer galaxies.} We used both flux-limited tracers ($i\leq
22.5$) and volume-limited tracers. For the latter, we imposed three
different luminosity limits given by $M_B \leq M_{lim}-Qz$ and
$M_{lim}=-19.9, -20.4, \text{and} -20.9$ to define tracer samples complete down
to $z= 0.8, 0.9,\text{and } 1.0$, respectively. We used $Q=1$ to account for the
evolution of the characteristic luminosity of the galaxy luminosity
function (see e.g. \citealp{zucca09} and
\citealp{kovac2010_density}). By using the flux-limited tracers,
$\rho$ can be computed on much smaller scales than with volume-limited
tracers (which are sparser by definition). In contrast, by using the
volume-limited tracers, we can compute $\rho$ with the same tracer
population at all redshifts.

{\bf 2) Smoothing filter.} For the VIPERS sample, we computed $\rho$
with different smoothing filters: i) cylinders with a half-length of
$1000 \kms$ and radius corresponding to the distance to the fifth,
tenth, and twentieth n.n. (we call these radii $R_{5th}$,
$R_{10th}$, and $R_{20th}$); ii) cylinders with a half-length of $1000
\kms$ and radius $R= 2, 3, 4, 5, \text{and }8 \mpcoh$ (radii $R_{2c}$, $R_{3c}$,
$R_{4c}$, $R_{5c}$, and $R_{8c}$); and iii) spheres with radius $R=5$ and
$8 \mpcoh$ as done in \cite{cucciati14} (radii $R_{5s}$ and
$R_{8s}$). A fixed-scale $R$ allows us to study the entire range of
local densities on the same scale. In contrast, when we use an
adaptive $R,$ we are able to reach much smaller scales in high-density
regions (but larger scales in low-density regions). A cylindrical
filter, being elongated along the line of sight, is better suited for redshift space, like in our case: in this way, we can
better account for the peculiar velocities of galaxies in high
densities while keeping a relatively small scale on the plane of the
sky. In the case of cylinders, the radius $R$ is computed in the
R.A.-Dec plane from the cylinder centre, and for the adaptive radius
the $n^{th}$ n.n. is computed in 2D on the plane of the sky,
considering all galaxies within $\pm1000 \kms$ to be at the same
redshift. For all the tracers within the filter, we set $F(R)=1/ (\pi
R^2)$, and $\rho({\bf r})$ has the dimensions of a surface
density. For the spherical filter, tracer counts are made in 3D
comoving space within a sphere of radius $R$, and we set $F(R)= 1/
(4/3 \pi R^3)$. In this case, $\rho({\bf r})$ has the dimensions of a
volume density.

{\bf 3) Cell position.}  The aim of this work is to study how the
local environment around galaxies affects their evolution, therefore we
centred the cells on our galaxies. Tracer galaxies at the centre of a
smoothing element are included in the count.

{\bf 4) Weights.} The function $\phi$ in Eq.~\ref{rho_eq} normally
takes into account the survey selection function.  In the case of
VIPERS, $\phi$ should include the weights CSR, TSR, and SSR discussed
in Sect.~\ref{data}. Alternatively, we can use $\phi=1$ for our
spectroscopic sample, but we also include in the tracer sample the
galaxies for which we only have a photometric redshift. In this way, our
tracer sample will be complete in the studied volume. We followed this
second approach, and we minimised the effects of the large photometric
redshift error by modifying the PDF of $z_p$ using the `ZADE' method (see
Appendix~\ref{zade}). We refer to \cite{cucciati14} for a
detailed comparison of the two approaches. Briefly, the ZADE method
allows us a density reconstruction with much smaller random errors.

\begin{figure} \centering
\includegraphics[width=\hsize]{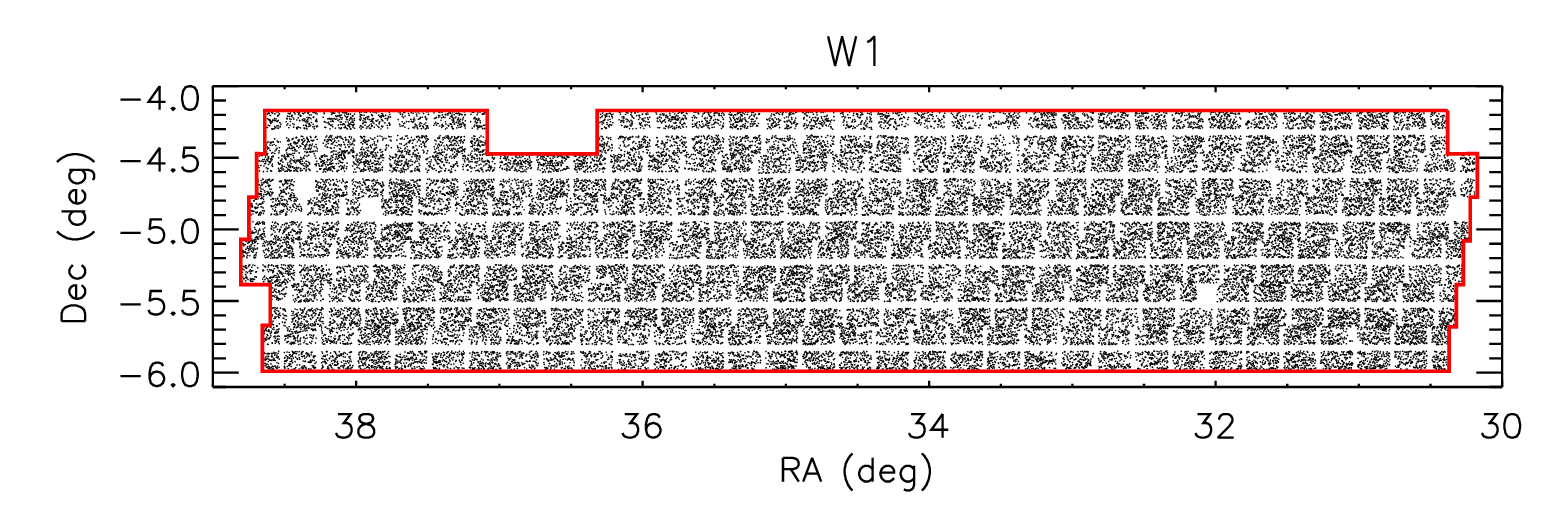}
\includegraphics[width=\hsize]{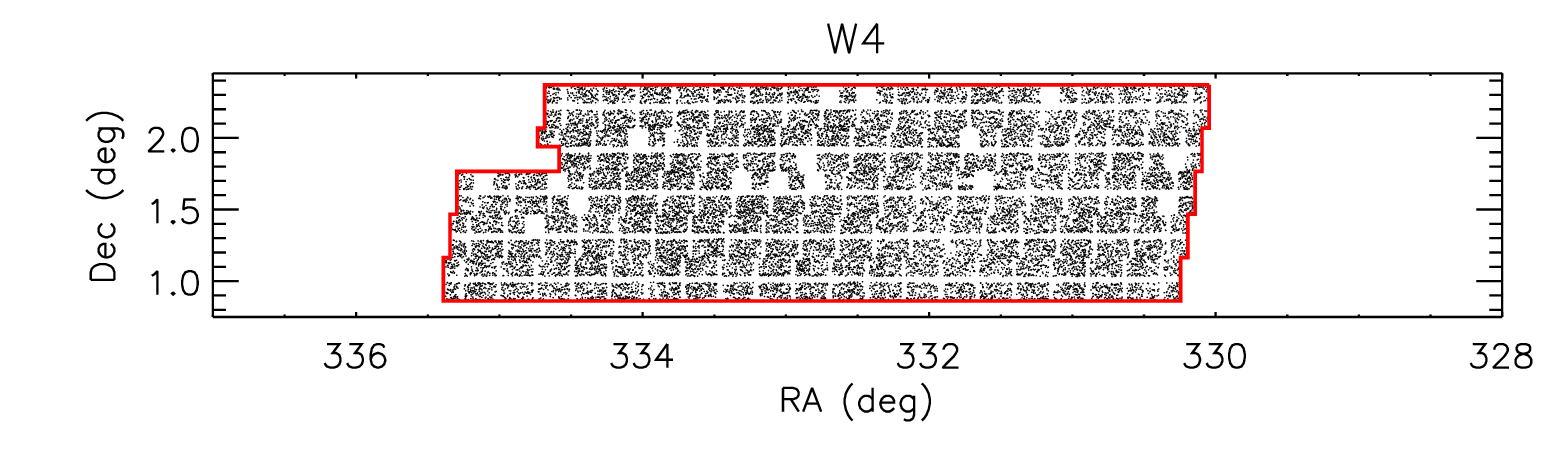}
\caption{R.A.-Dec distribution of secure redshift galaxies in W1 (top)
  and W4 (below). The red thick line in each panel is the `field
  boundary' that we consider in this work. The cross-like pattern of
  void regions is due to the characteristic footprint of the VIMOS
  instrument. Rectangular empty regions are missing quadrants
that   have been discarded due to too poor observational conditions or
  technical problems.}
\label{fields_fig} 
\end{figure}

{\bf 5) Gaps and survey boundaries.} As extensively discussed in
\cite{cucciati14}, we fill the cross-like pattern (`gaps') typical of
VIMOS observations using the galaxies with photometric redshifts, to
which we apply the ZADE method. The same holds for the empty regions
corresponding to missing quadrants. In contrast, we take into account
the `field boundaries' (the true limits in R.A. and Dec of the
surveyed area, in red in Fig.~\ref{fields_fig}) as follows: when a
cell falls partially outside the field boundaries, we divide
$\rho({\bf r})$ by the fraction of the filter that falls within the
boundaries before computing $\delta({\bf r})$ (see
\citealp{cucciati2006}).

{\bf 6) Mean density.}  We compute the mean density $\langle \rho({\bf
  r}(z)) \rangle$ by smoothing the VIPERS $n(z)$ with a statistical
approach based on the $V_{max}$ method. Full details are given in
\cite{kovac2010_density} (but see also \citealp{marulli13} and
\citealp{delatorre13a}). In the case of the spherical filter (which is
based on a 3D distance), we then compute $\langle \rho({\bf r}(z))
\rangle$ dividing the $n(z)$ by the volume of the survey. In the case
of cylindrical filters, we obtain $\langle \rho({\bf r}(z)) \rangle$
by integrating $n(z)$ in a redshift range of $\pm 1000 \kms$ centred
on the redshift of each given galaxy and by dividing the result by the
survey area.

\subsection{ZADE}\label{zade}

We use a modified version of the ZADE approach described in
\cite{kovac2010_density}. All the details of the performance of ZADE
in a VIPERS-like survey with respect to other methods are given in
\cite{cucciati14}. \cite{cucciati14} tested the performance of the
method using spherical cells. Since the performance is not expected to
depend on the filter shape, we adopted the same method here.

The method is applied to each galaxy for which we have only the
photometric redshift, and it can be described as follows.  For each of
these galaxies, we keep its position in the sky (R.A. and Dec.), and
we collapse the probability distribution function (PDF) of its
photometric redshift $z_P$ on several probability peaks along the
l.o.s.. The redshifts of these peaks are determined by the peaks of
the $n(z)$ of the spectroscopic galaxies falling in a cylinder centred
on the position of the given photometric galaxy (R.A., Dec.  and
$z_p$), with radius $R_{ZADE}=5\mpcoh$ and half-length equal to
$3\sigma_{zp}$. The weights $w_{ZADE}$ assigned to these peaks are
given by the product of the PDF by $n(z)$, normalised to unity.

By applying this method, the summation in Eq.~\ref{rho_eq} runs over
all the spectroscopic galaxies setting $\phi=1$ and over all the
peaks setting $\phi$ equal to each given $w_{ZADE}$. Counts are
performed only among the galaxies that satisfy the selection criteria
to be considered as tracers for both spectroscopic and photometric
galaxies.

\begin{figure} \centering
\includegraphics[width=4.3cm]{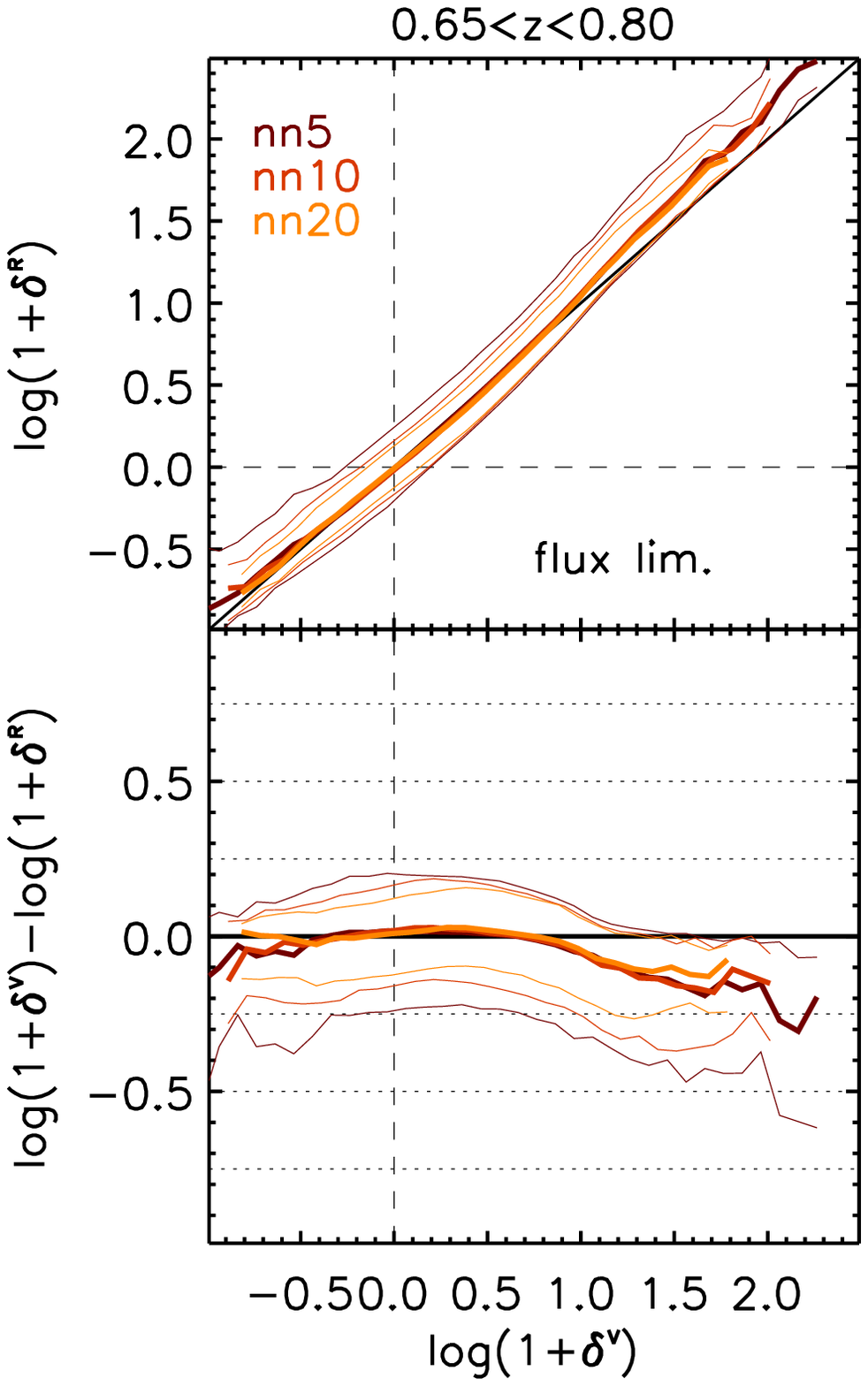}
\includegraphics[width=4.3cm]{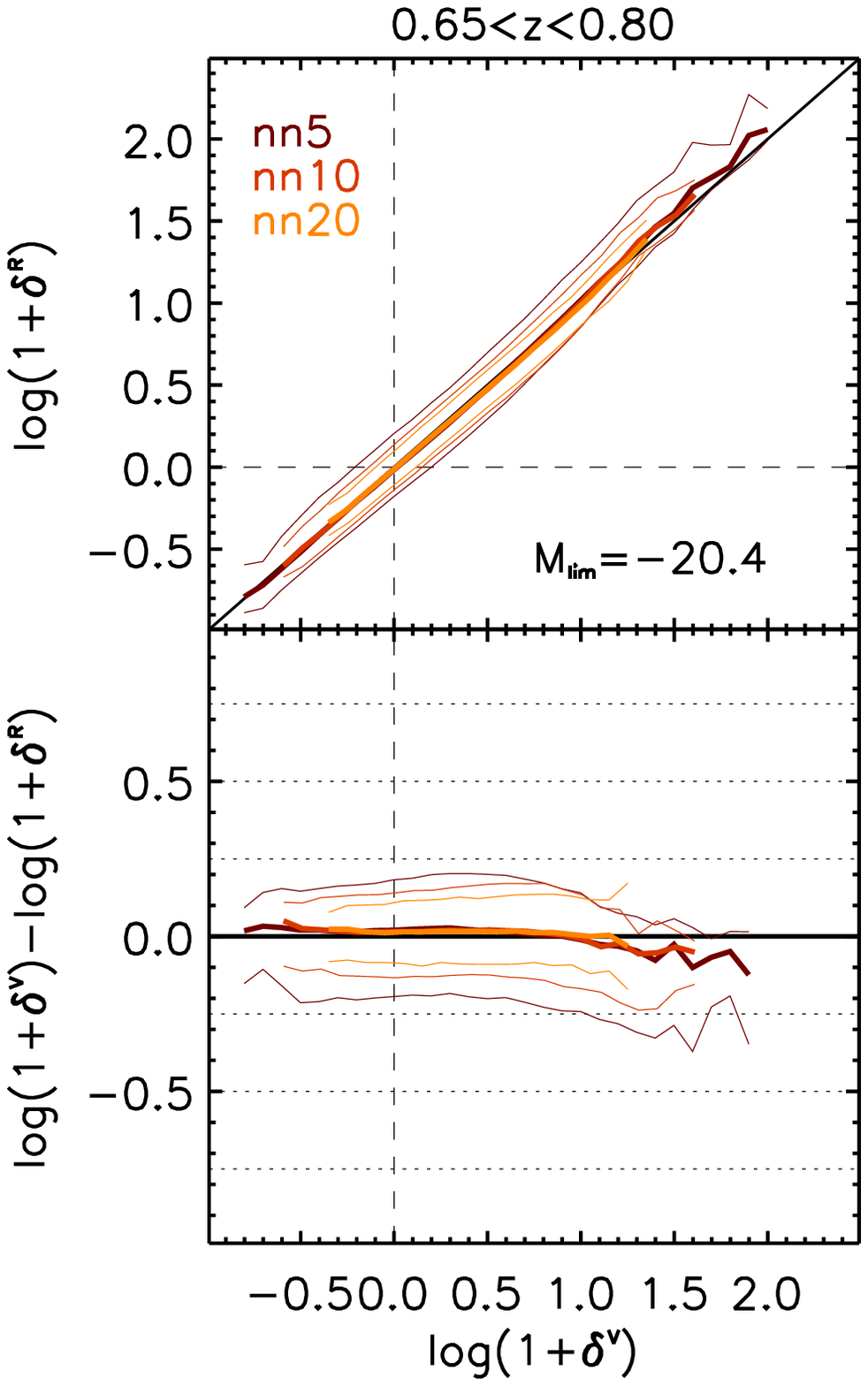}
\caption{Comparison of $\delta^R$ and $\delta^V$ on a galaxy-by-galaxy
  basis.  Results are shown for the cylindrical filter with radius
  corresponding to the fifth, tenth, and twentieth n.n. for the
  redshift bin $0.65<z<0.8$. The density is computed using flux-limited tracers (left) and volume-limited tracers with
  $M_{lim}=-20.4$ (right). $x$-axis: density contrast in the \vmop;
  $y$-axis, top panel: density contrast in the \rmop; $y$-axis, bottom
  panel: difference of the logarithms. The thick lines are the median
  value of the quantity displayed on the $y$-axis in each $x$-axis
  bin. Thin lines represent the sixteenth  and eightyfourth percentiles of its
  distribution.  The solid black line in the top panels is the
  one-to-one line, and the horizontal lines in the bottom panels are
  for reference.}
\label{dens_test_nn} 
\end{figure}

\begin{figure} \centering
\includegraphics[width=4.3cm]{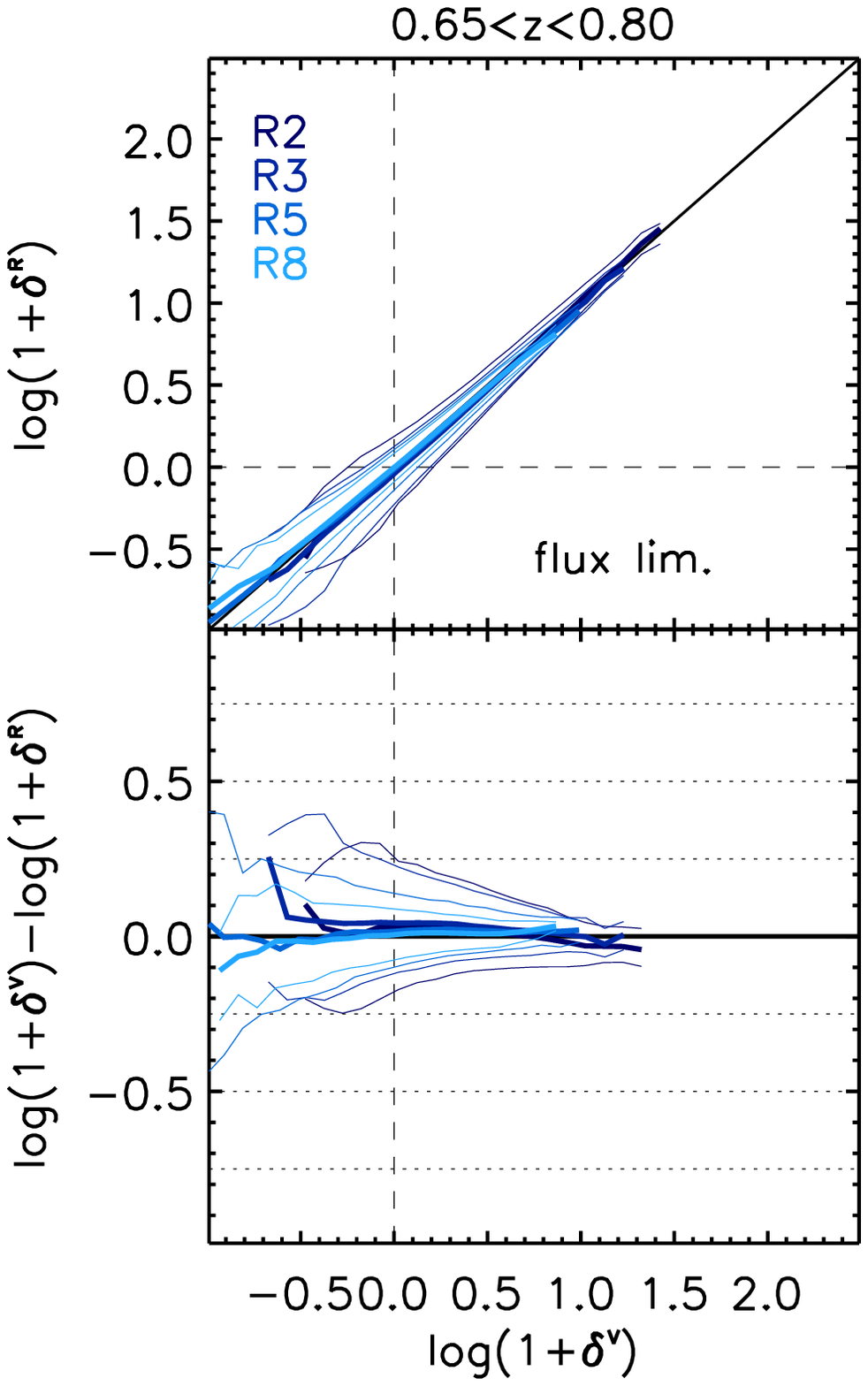}
\includegraphics[width=4.3cm]{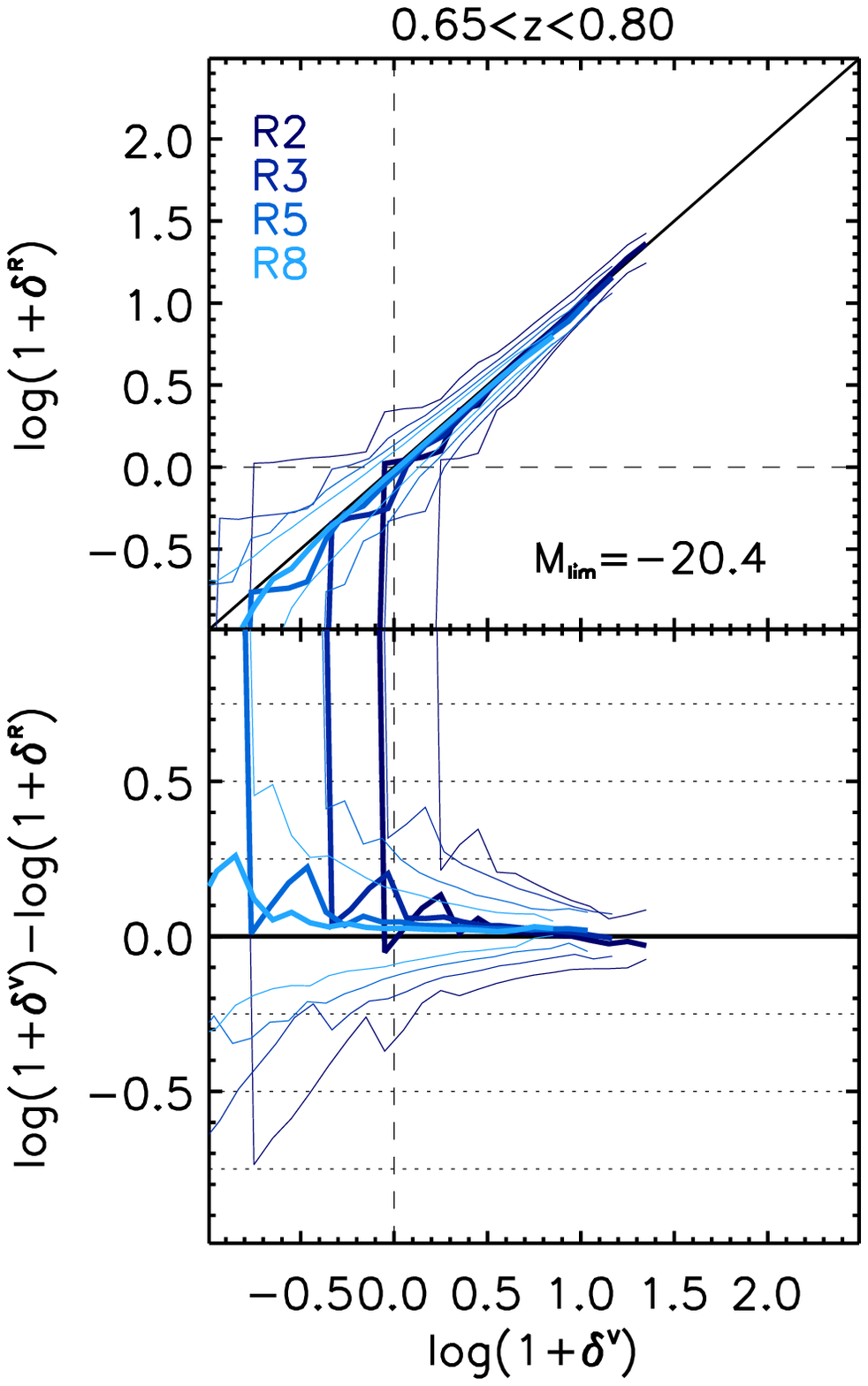}
\caption{As in Fig.\ref{dens_test_nn}, but for cylinders with fixed
  radius. }
\label{dens_test_R} 
\end{figure}

\subsection{Reliability of the density reconstruction}\label{robustness_dens_app}

We compute the density contrast $\delta$ in the \rmops and \vmops
($\delta^{R}$ and $\delta^{V}$, respectively) using Eqs.\ref{delta_eq}
and \ref{rho_eq}. In the \rmops the filter is centred on all the
galaxies, and the summation in Eq.~\ref{rho_eq} runs on all the
tracers, each with $\phi=1$. In the \vmops we centre the filters only
around spectroscopic galaxies, and we apply the ZADE method. This way,
the summation in Eq.~\ref{rho_eq} runs over all the spectroscopic
galaxies with $\phi=1$, and over all the ZADE peaks, for each setting
$\phi=w_{ZADE}$. Then we match each \rmos with the spectroscopic
catalogue of its corresponding \vmo, to have $\delta^{R}$ and
$\delta^{V}$ measured for the same set of galaxies\footnote{To assess
  the reliability of the VIPERS density field reconstruction, we used
  also a set of light cones derived with a HOD method as described in
  \cite{delatorre13a}, which are the same used in \cite{cucciati14}
  and D16. The results obtained with these HOD mock catalogues are
  very similar to the result described here, therefore we do not show them.}.

We compare $\delta^{R}$ and $\delta^{V}$in two ways: on a
galaxy-by-galaxy basis, to assess how well we can recover the density
absolute value, and according to their ranking, to assess how well we
can separate low- and high-density regions. We performed the
comparison in three redshift bins, the same used for the scientific
analysis of this paper: $0.51<z\leq0.65$, $0.65<z\leq0.8$, and
$0.8<z\leq0.9$. We note that the results of this comparison depend
very mildly on redshift, therefore we only show the results for the
central redshift bin here.

Figures~\ref{dens_test_nn} and~\ref{dens_test_R} show the
galaxy-by-galaxy comparison. We show only the cases of the
flux-limited sample and the brightest volume-limited sample, as they
represent the two most extreme types of tracers.  We note that for fixed
radius and volume-limited tracers (right panels in
Fig.~\ref{dens_test_R}) the sharp discontinuity to infinity is an
artefact corresponding to the case of zero counts. We see that

\begin{itemize}

\item[-] for both the adaptive and fixed radii, the systematic error
  is always close to zero except for the highest densities in the case
  of flux-limited and adaptive radius, where we underestimate the real
  density by $\sim 30\%$.

\item[-] The random error does not depend on density for the adaptive
  radii, and it varies between $\sim40\%$ and $\sim15\%$ from
  $R_{5th}$ to $R_{20th}$; in contrast, it depends on density for the
  fixed radii, dropping from $>40\%$ to $\sim 10\%$ when moving from
  low to high density. This different behaviour is due to fixed or
  varying number of galaxies within the cylinders in the case of the
  adaptive or fixed radius, respectively.

\item[-] In all cases, the random error is smaller for larger radii,
  and neither the systematic nor the random error seems to depend on
  redshift.
\end{itemize}

We note that one of the reasons why the high densities are
underestimated is that VIPERS is a single-pass survey, so that
close pairs
are more difficult to target (unless the two galaxies are so close
and at the same Dec. that they fall in the same slit). The use of
ZADE partially mitigates this loss because it allows us to use all the
not-targeted galaxies (although with a larger error on the position
along the l.o.s.).

We also performed a less demanding test: we compared the
$\delta^{R}$ and $\delta^{V}$ values according to their ranking, and
not to their value. We divided $\delta^{R}$ and $\delta^{V}$ into
quartiles, and we focused on the first quartile (the
lowest densities, `LD'), and the last quartile (the highest densities,
`HD').

We call $N^V_{i}$ ($N^R_{i}$) the number of galaxies falling in the
percentile $i$ of the $\delta^V$ ($\delta^R$) distribution, with $i$
equal to LD or HD.  Also, we call $N_{i,j}^{V,R}$ the number of
galaxies that fall in the percentile $i$ of $\delta^V$ and in the
percentile $j$ of $\delta^R$, with $j$ also equal to LD or HD. We then
define the completeness and contamination of the percentile $i$ as
\begin{equation} \displaystyle
\text{completeness} = N_{i,i}^{V,R} / N^R_i
\label{comp_eq} 
\end{equation}
\begin{equation} \displaystyle
\text{contamination} = N_{i,j}^{V,R} / N^V_i  \qquad \text{(with $i \neq j$)}
\label{cont_eq} 
.\end{equation}

In practice, the completeness expresses the fraction of galaxies that
are placed in the correct percentile, while the contamination
indicates which fraction of galaxies belonging to the LD(/HD)
percentile of $\delta^V$ distribution come from the opposite
percentile HD(/LD) of the original $\delta^R$ distribution. The best
result would be a completeness of 100\% and a contamination equal to
zero.

The values of completeness and contamination for the different types of
filters and radii are shown in Table \ref{PC_tab}.  We find that the
values of completeness and contamination mirror the results shown in
Fig.~\ref{dens_test_nn} and \ref{dens_test_R}, and they mainly depend
on the random error shown in these figures. The relatively low random
errors allows us to obtain quite high levels of completeness (almost
always above 75\%), and basically zero contamination. Moreover, for
cylinders with a fixed radius, the completeness is higher in HD than in LD,
corresponding to the larger random error in LD.

We conclude that we can distinguish with confidence between the lowest and
highest density regions, each time selecting highly complete samples
with very low contamination from the opposite environment. This is
particularly important for the scientific analysis in this paper.

\begin{table} 
  \caption{Completeness and contamination for LD and HD quartiles for 
    different types of filter and different radii and tracers. Values represent the average 
    over the three redshift bins when the spread among the three values was 
    below 3\%, otherwise the minimum and maximum values are given.}  
\label{PC_tab} 
\centering 
\begin{tabular}{l c c c c } 
  \hline   
  \hline   
  Filter & \multicolumn{2}{c}{Completeness (\%)} & \multicolumn{2}{c}{Contamination (\%)} \\
  &  LD &  HD  &  LD  & HD  \\    
  \hline                                                                   
\multicolumn{5}{c}{Volume-limited tracers ($M_{lim}=-20.4$)} \\
  $R_{5th}$   & 73.8  &  74.1   & $<1$  &  $<1$  \\
  $R_{10th}$  & 75.9  &  76.7   & $<1$  &  $<1$  \\ 
  $R_{20th}$  & 77.8  &  78.4   & $<1$  & $<1$  \\ 
  $R_{2c}$    & 68.8  &  74.0    & $<1$  & $<1$   \\ 
  $R_{3c}$    & 70.5  &  76.2    & $<1$  & $<1$   \\ 
  $R_{5c}$    & 75.9  &  79.8    & $<1$  & $<1$   \\ 
  $R_{8c}$    & 80.2  &  82.7    & $<1$  &  $<1$  \\ 
  \hline                                                                   
\multicolumn{5}{c}{Flux-limited tracers} \\
  $R_{5th}$   & 74.1  & 74.9   & 1.1  &  $<1$  \\
  $R_{10th}$  & 76.6  & 78.7    & $<1$ &   $<1$  \\ 
  $R_{20th}$  & 78.8  & 81.3    & $<1$  &  $<1$  \\ 
  $R_{2c}$  & 70-75 & 76-82    & $<1$  & $<1$   \\ 
  $R_{3c}$  & 74-78 & 78-84    & $<1$  & $<1$   \\ 
  $R_{5c}$  & 78-83 & 81-87    & $<1$  & $<1$   \\ 
  $R_{8c}$  & 82-87 & 84-88    & $<1$  & $<1$   \\ 
  \hline 
\end{tabular} 
\end{table}

\section{Density field in the PDR-1}\label{PDR1_vs_PDR2}

\begin{figure} \centering
\includegraphics[width=4.3cm]{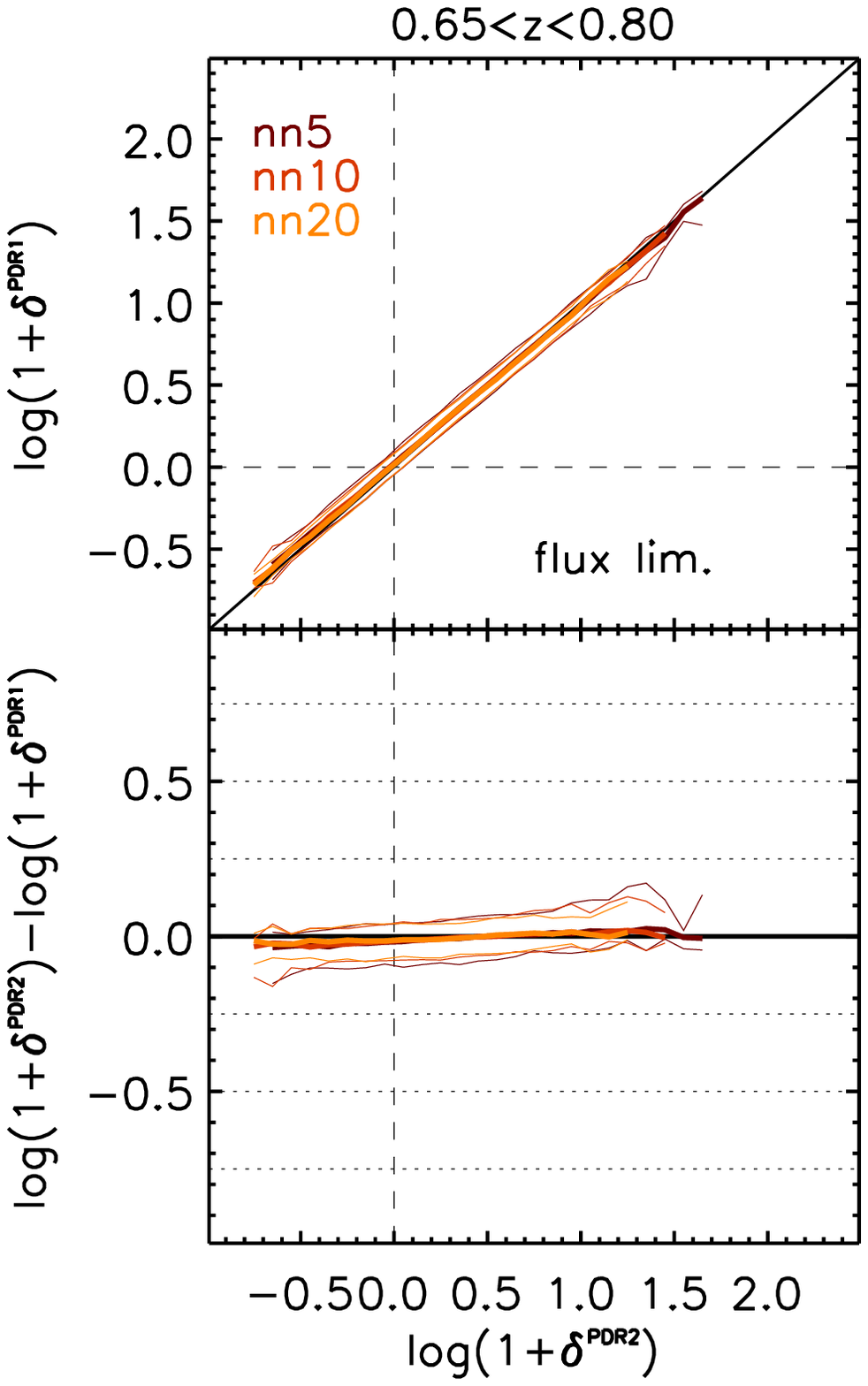}
\includegraphics[width=4.3cm]{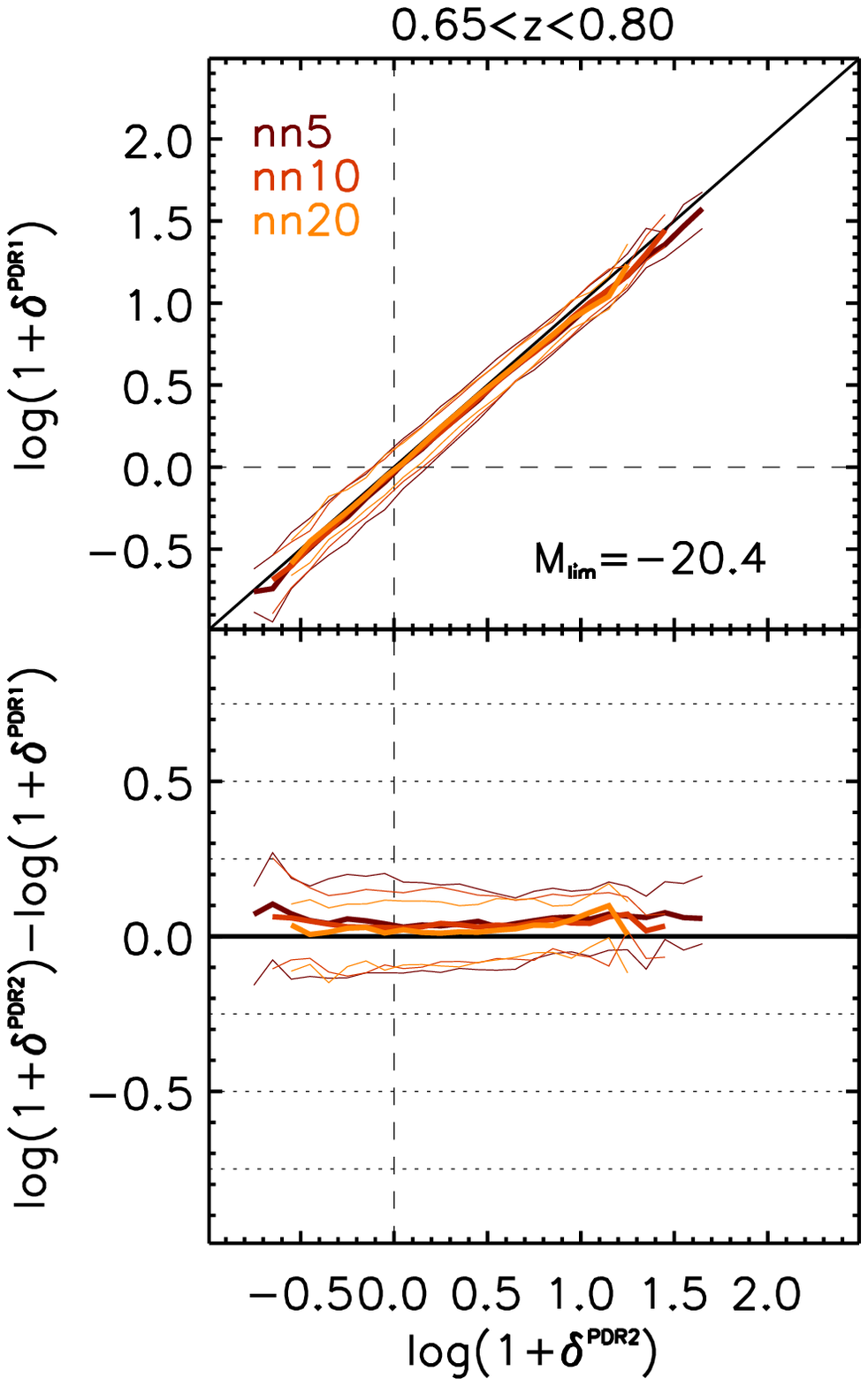}
\caption{As in Fig.~\ref{dens_test_nn}, but here we compare the
  density field in the VIPERS PDR-1 sample and the density field used
  in this paper (for simplicity indicated as PDR-2 in the labels).}
\label{dens_PDR1_PDR2} 
\end{figure}

In this appendix we compare the density field computed for the VIPERS
PDR-1 sample, used in D16, with the density field that we use in this paper. The
comparison is shown in Fig.~\ref{dens_PDR1_PDR2} for the galaxies in
common to both the samples. We only show one redshift bin because the
results in the two other bins are very similar. The
systematic difference is almost zero, and the scatter around the mean
difference is $\lesssim10\%$ for the flux-limited tracers and
$\lesssim25\%$ for the volume-limited tracers. In both cases, it is
smaller than the random error in the density field reconstruction due
to the VIPERS observational strategy (Fig.~\ref{dens_test_nn}). We
also remark that the systematic and random errors are both smaller for
larger radii.

\section{ sSFR distribution in the model}\label{sSFR_model}

In Fig.~\ref{sSFR_distrib} we have shown that the sSFR distribution in
the model is different from the data distribution in several
aspects. We verified whether these differences could be due to the
lack of measurement errors in the model sSFR by adding an error to
the model stellar mass or to the model SFR or to both of them,
extracted from a Gaussian distribution with $\sigma=0.25$ (dex).

Firstly, we analysed the tail of high sSFR, which is missing in the
model. By adding the error only to the stellar mass, the tail of high
sSFR shrinks even further, as expected. In fact, if we add a Gaussian
error to the stellar mass, given the shape of the GSMF, galaxy masses
are preferentially boosted. We therefore expect the sSFR to
decrease. Instead, if we also add an error to the SFR, the high-sSFR
tail of the sSFR distribution tends to increase as we increase the SFR
error. In this way, we can recover the observed tail of high sSFR.  We
remark that the model and observed sSFR tails are increasingly similar
at higher redshift, so that we would need to use different SFR errors at
different redshifts to match the models to the observations. Moreover,
if we consider the two errors on stellar mass and on SFR to be
correlated, the overall shape of the model sSFR is preserved (e.g.,
the sort of plateau at $\log({\rm sSFR})\sim-11$), while when
we use
uncorrelated errors, the sSFR distribution becomes smoother, assuming
the shape of a single-peaked skewed distribution.

Secondly, the data sSFR distribution features a valley at $\log({\rm
  sSFR})\sim-10.8,$ whereas the model predicts a plateau.  We are
unable to reproduce this valley by adding an error to the model
stellar mass and SFR. Correlated errors do not alter the shape of this
plateau, while uncorrelated errors removed the point of inflection
altogether.

Although by adding these measurement errors we obtain a better
agreement between the model and the observed sSFR distributions, we
decided not to include them in the model sSFR because
modelling them accurately is beyond the scope of this paper.

Given the different shape of the model and observed sSFR
distributions, we do not use the thresholds $\log({\rm sSFR})<-11.2$
and $\log({\rm sSFR})>-10.8$ (used for the VIPERS data, see
Sect.~\ref{act_pass}) to define passive and active galaxies in the
model. Instead, in each redshift bin, considering only galaxies above
the mass limit, we computed the fraction of passive and active
galaxies in the data (defined using the sSFR thresholds).  We used the
same fractions in the model, starting to count galaxies from the two
tails of the sSFR distribution. The comparison between the data and
the model should be more meaningful when we select the same `extremes'
of the sSFR distribution. In the data, the fraction of passive
galaxies ranges from 35\% to 44\% from the lowest to the highest
redshift bin, and the fraction of active galaxies ranges from 52\% to
40\%. In the model, these fractions of passive galaxies roughly
correspond to selecting all galaxies with $\log({\rm
  sSFR})<-11.66,-11.30,-11.23$ from the lowest to the highest
redshift. The fractions of active galaxies correspond to a selection
given by $\log({\rm sSFR})>-10.7,-10.5,\text{and}-10.4$.

We note that we compute these fractions considering only the galaxies
above the stellar mass limit in each redshift bin.  This is because,
as shown in \cite{davidzon13}, the model GSMFs are more similar to
those in VIPERS for $\log(\mathcal{M}/\mathcal{M}_\odot)\gtrsim10.7$.
This allows us to work within a stellar mass range where our model and
the data are in better agreement.

\end{document}